\documentclass[final,5p,times,twocolumn]{elsarticle}

\usepackage{amssymb}
\usepackage{blindtext}
\usepackage{graphicx}
\usepackage{subcaption}
\usepackage{amsfonts,bm}
\usepackage[cmex10]{amsmath}
\usepackage[colorlinks=true, allcolors=blue]{hyperref}
\usepackage{caption}
\usepackage{lineno}
\usepackage{stackrel}

\usepackage[dvipsnames]{xcolor}
\usepackage[normalem]{ulem} %provides strike-through

\newcommand{\CR}{\ensuremath{\mathbb{C}^{\R}}}

\DeclareFontFamily{U}{mathx}{\hyphenchar\font45}
\DeclareFontShape{U}{mathx}{m}{n}{
      <5> <6> <7> <8> <9> <10>
      <10.95> <12> <14.4> <17.28> <20.74> <24.88>
      mathx10
      }{}
\DeclareSymbolFont{mathx}{U}{mathx}{m}{n}
\DeclareFontSubstitution{U}{mathx}{m}{n}
\DeclareMathAccent{\widecheck}{0}{mathx}{"71}

\newcommand{\Frac}[2]{{{#1}/{#2}}}  % an "inert" form of \frac
\newcommand{\bias}{\operatorname{bias}}
\newcommand{\MSE}{\operatorname{MSE}}

\newcommand{\beq}{\begin{equation}}
\newcommand{\eeq}{\end{equation}}
\newcommand{\beqan}{\begin{eqnarray*}}
\newcommand{\eeqan}{\end{eqnarray*}}

\newcommand{\eqlabel}[1]{ \stackrel{(#1)}{=} }

\newcommand{\approxlabel}[1]{ \stackrel{(#1)}{\approx} }

\newcommand{\openCase}  {\left\{ \begin{array}{@{\,}ll}}
\newcommand{\openCasell}{\left\{ \begin{array}{@{\,}ll}}
\newcommand{\openCasecl}{\left\{ \begin{array}{@{\,}cl}}
\newcommand{\openCaserl}{\left\{ \begin{array}{@{\,}rl}}
\newcommand{\openCaseTablell}{\left\{ \begin{array}{@{}ll}}
\newcommand{\openCaseTablecl}{\left\{ \begin{array}{@{}cl}}
\newcommand{\openCaseTablerl}{\left\{ \begin{array}{@{}rl}}
\newcommand{\closeCase} {\end{array} \right.}

\DeclareMathOperator*{\argmax}{arg\,max}

\renewcommand{\P}[1]{\mathrm{P}\left({#1}\right)} % Probability
\def\smid{\,|\,}  % Like \mid, but with less space; use for conditioning
\def\sMid{\,;\,}  % Version of \smid for non-Bayesian case
\newcommand{\E}[1]{\mathrm{E}\!\left[\,{#1}\,\right]}   % Expectation
\newcommand{\iE}[1]{\mathrm{E}[\,{#1}\,]}   % Expectation classical (inline)
\newcommand{\var}[1]{\mathrm{var}\!\left({#1}\right)}   % Variance
\def\CR{Cram{\'e}r--Rao}

\def\etaBaseline{\hat{\eta}_{\rm baseline}}
\def\etaOracle{\hat{\eta}_{\rm oracle}}
\def\etaTR{\hat{\eta}_{\rm TR}}
\def\utilde{{\widetilde{u}}}
\def\Utilde{{\widetilde{U}}}
\def\MSE{\widehat{\mbox{\footnotesize MSE}}}
\def\MSElower{\mbox{\footnotesize MSE}^-}
\def\MSEupper{\mbox{\footnotesize MSE}^+}
\def\MSEcaption{\widehat{\mbox{\scriptsize MSE}}}
\def\MSElowercaption{\mbox{\scriptsize MSE}^-}
\def\MSEuppercaption{\mbox{\scriptsize MSE}^+}
\DeclareMathOperator{\Poisson}{Poisson}

\begin{document}

\begin{frontmatter}
\title{Source Shot Noise Mitigation in Focused Ion Beam Microscopy by \\
Time-Resolved Measurement
}

\author[label1]{Minxu Peng}
\address[label1]{Department of Electrical and Computer Engineering, Boston University, Boston, MA, 02215, US}

\author[label1]{John Murray-Bruce}

\author[label2]{Karl K. Berggren}
\address[label2]{Department of Electrical Engineering and Computer Science, Massachusetts Institute of Technology, Cambridge, MA 02139, US}

\author[label1]{Vivek K Goyal\corref{cor1}}
\ead{v.goyal@ieee.org}

\cortext[cor1]{Corresponding author.}

\begin{abstract}
Focused ion beam (FIB) microscopy suffers from source shot noise~--
random variation in the number of incident ions in any fixed dwell time~--
along with random variation in the number of detected secondary electrons
per incident ion.
This multiplicity of sources of randomness increases the variance of the
measurements and thus worsens the trade-off between incident ion dose and
image accuracy.
Time-resolved sensing combined with maximum likelihood estimation from the
resulting sets of measurements greatly reduces the effect of source shot noise.
Through Fisher information analysis and Monte Carlo simulations,
the reduction in mean-squared error or reduction in required dose is shown to
be by a factor approximately equal to the secondary electron yield.
Experiments with a helium ion microscope (HIM) are consistent with the
analyses and suggest accuracy improvement for a fixed source dose,
or reduced source dose for a desired imaging accuracy,
by a factor of about 3\@.
\end{abstract}


\end{frontmatter}

\thispagestyle{empty}

\section{Introduction}
State-of-the-art techniques for imaging the structure of a sample at near-atomic resolution
depend on the use of microscopes that scan the sample with a focused beam of particles.
For instance,
a focused electron beam is employed in scanning electron microscopy (SEM) \cite{McMullan1995},
laser beams in confocal laser-scanning microscopy \cite{Minsky1988} and two-photon laser-scanning fluorescence microscopy \cite{Denk1990},
and helium ion beams in helium ion microscopy (HIM) \cite{Ward2006}.
A fundamental goal with these technologies is to 
aim to produce the best image quality for a given number of incident particles.
This is especially relevant when each incident particle appreciably damages the sample;
thus, we henceforth concentrate on
HIM\@.

Focused ion beam (FIB) imaging methods have randomness in the number of incident particles (the \emph{source shot noise}) and in the influence of each incident particle on the instrument measurement.
The goal of the imaging is to infer properties of the sample that are
revealed through the number of detected secondary electrons (SEs)
per incident ion,
and the source shot noise is
detrimental to this effort because it is unrelated to the sample.
Presumably, we would prefer to have a precisely known number of incident ions.

The main idea of this work is that time-resolved measurement of SEs
can be used to mitigate the effect of source shot noise.
In certain limiting cases, we can completely eliminate the effect of source
shot noise, producing estimation performance equivalent to a deterministic
incident ion beam.
More importantly, for parameters that reasonably model HIM, 
the improvement is substantial and validated by both simulations
and experiments.

The key technical result is an analysis of Fisher information gain
and the consequent observation that information gain per incident particle
is maximized when the number of incident particles per measurement trial is low
(e.g., from a combination of low dwell time and low beam current).
Based on this, we advocate for combining information from
multiple low-intensity acquisitions,
which we term \emph{time-resolved} (TR) measurement.
We first presented the TR measurement concept for FIB microscopy in~\cite{Peng:18}.

\subsection{Background}
The first image of a solid sample based on secondary electrons emitted in response to an electron beam scanner was produced by Knoll in 1935, inspiring the development of a dedicated SEM \cite{McMullan1995}.
Ever since their development, SEMs have been ubiquitous in both research and industrial imaging, as well as in nanometerological applications \cite{spence2007diffractive}.
Building upon decades of research in focused ion beam microscopy, the first commercial HIM was introduced in 2006 \cite{Ward2006,Economou2012}, with the promise of producing images with sub-nanometer resolution~\cite{joens2013helium} and reduced charging of the sample, when compared with SEM\@.
However just like SEM, HIM uses a focused particle beam to produce lateral spatial resolution in a ballistic configuration \cite{Ward2006}. Both material composition (e.g., atomic number) and shape (topographic yield variations common to SEM as well) contribute to the number of SEs dislodged from the specimen~\cite{Ramachandra2009a}.
These properties, along with improved diffraction-limited imaging resolution
and reduced sample charging, have enabled superior imaging of insulators without the need for metal coating.
Hence HIM is an important imaging technology for semiconductor and nanofabrication research~\cite{schurmann2015helium}.

Notwithstanding the progress in the pursuit of ultra-high resolution, these imaging
technologies all have the disadvantage of causing damage to the sample through
sputtering~\cite{castaldo2011simulation,castaldo2009influence, orloff1996fundamental}. Whilst sample damage can have
especially severe impact on biological samples, it also occurs for many other
types of materials. It is thus recognized and modeled as a fundamental limit to imaging with focused beams.
With the helium ion being 7300 times more massive than the electron, mitigating sample damage in HIM is paramount. One possible approach is imaging
using lower ion doses but at the cost of image lower quality~\cite{castaldo2009influence}.
Consequently, studies analyzing the extent of beam
damage and establishing safe imaging dose have appeared~\cite{fox2013helium, livengood2009subsurface}.

\subsection{Main contributions}
\begin{itemize}
\item Introduction of time-resolved measurement as a mechanism for
      mitigation of source shot noise in FIB microscopy.
\item Introduction of mathematical models for FIB microscopy with a
      Poisson number of incident ions, Poisson number of SEs
      per incident ion, and direct or indirect detection of the SEs.
\item Quantitative analyses of Fisher information gain for the above
      models, including comprehensible expressions for the low- and
      high-dose limits for the direct-detection model.
\item Experimental demonstration of the use of time-resolved measurement
      using data from a Zeiss ORION NanoFab HIM\@.
      Despite a lack of ground truth, evidence of improvements over
      conventional image formation is compelling.
\end{itemize}

\subsection{Outline}
In Section~\ref{sec:Model_methods},
we present our baseline measurement model and basic analyses of this model.
These analyses provide the foundations for our development,
in Section~\ref{sec:TR}, of the advantage provided by dividing any fixed ion dose
into small doses through time-resolved measurement.
We present both abstract numerical results and image simulations.
Inspired by the indirect detection of SEs in current HIM instruments,
Section~\ref{sec:hierarchical_models} introduces suitable models and studies the theoretical improvement factors from time-resolved measurement.
Section~\ref{sec:HIM-results}
presents experimental results using data from a Zeiss HIM\@.

\section{Single measurement: model and analyses}
\label{sec:Model_methods}
Two main components enable FIB imaging: a stable source to generate the FIB and a detector to measure the number of SEs leaving the sample's surface.
Due to ion--sample interaction, SEs become excited and dislodged from the sample's surface~\cite{cazaux2010calculated}, accelerating towards the SE detector.
Imaging is achieved by raster scanning the ion beam with some fixed dwell time per pixel.
For each pixel, detected SEs are mapped to a grayscale level, hence producing an image of the sample.

During the acquisition process, for any fixed dwell time there is randomness in the number of ions reaching the sample. In addition, for each ion that interacts with the sample, there is randomness in the number of emitted SEs.
In this section, we discuss a ``Poisson--Poisson'' model in which both the numbers of ions and the numbers of SEs induced by each ion follow Poisson distributions.
With this model,
the estimability of mean SE yield is amenable to theoretical analysis
through Fisher information.
The analyses of this section are used to support the use of time-resolved
measurement in Section~\ref{sec:TR}, and 
richer models are considered in Section~\ref{sec:hierarchical_models}.
All the analyses and methods of this paper are applied separately
for each micrograph pixel, so we do not include any pixel indexing.

\subsection{A Poisson--Poisson model for FIB imaging}
\label{subsec:PPModel}
In our abstraction,
an ion beam incident on the sample for a fixed dwell time
$t$ has ion arrivals following a Poisson process with rate $\Lambda$
per unit time.
Hence, the number of incident ions $M$ is a Poisson random variable
with mean
$\lambda = \Lambda t$.
Ion $i$ produces a number of SEs $X_i$ following a Poisson distribution with mean $\eta$.
Since emitted SEs travel a very short distance before being
captured by the SE detector, we model the
SE detections as instantaneous and simultaneous.
The fundamental assumption is that the delay before SE detection is much
less than a typical ion interarrival time;
this places some upper limit on the ion beam currents at which our model
is reasonable.

The goal is to produce an estimate of $\eta$ from
the total detected SEs
\begin{equation}\label{equ:Y-as-sum}
Y = \sum_{i=1}^M X_i,
\end{equation}
with $\lambda$ known.
Notice that $Y$ is a sum of $M$ independent Poisson random variables
where the unknown $M$ is itself also a Poisson random variable.
As shown in \ref{app:NeymanTypeA},
$Y$ is an example of a compound Poisson random variable;
specifically, it has the so-called \emph{Neyman Type~A} distribution~\cite{Neyman:39,Teich:81},
with probability mass function (PMF)
\begin{equation}\label{equ:neyman}
  P_{Y}(y \sMid \eta, \lambda)
    = \frac{e^{-\lambda} \eta^y}{y!}\sum\limits_{m = 0}^{\infty} \frac{(\lambda e^{-\eta})^m m^y}{m!},
\end{equation}
mean
\begin{equation}\label{eq:Y-mean}
\iE{Y} = \lambda \eta,
\end{equation}
and variance
\begin{equation}\label{eq:Y-variance}
\var{Y} = \lambda \eta + \lambda \eta^2.
\end{equation}
Ward et al.~\cite{Ward1991} demonstrated empirically that this is an accurate model
for numbers of detected SEs in an experimental setup involving a gallium ion beam.

\subsection{Baseline estimator}
\label{ssec:baseline}
It follows from \eqref{eq:Y-mean} that simple scaling,
\begin{equation}\label{eq:eta-baseline}
  \etaBaseline(Y) = \frac{Y}{\lambda},
\end{equation}
gives an unbiased
estimate of $\eta$. 
The mean-squared error (MSE) of this estimate,
\begin{align}
  \mathrm{MSE}(\etaBaseline)
  &= \E{ (\eta - \etaBaseline(Y))^2 } \nonumber \\
  &= \frac{\var{Y}}{\lambda^2}
   =  \frac{\eta(1+\eta)}{\lambda},
\label{eq:MSE-baseline}
\end{align}
thus follows from \eqref{eq:Y-variance}.
In imaging (in contrast to metrology), the scaling may be arbitrary;
thus, when every pixel has the same mean dose $\lambda$,
the SE counts can be used directly to form a reasonable image.

Assuming for the moment that $\lambda$ is an integer,
if the number of incident ions were deterministically $\lambda$,
the baseline estimator would be the sample mean of $\{X_i\}_{i=1}^\lambda$.
Furthermore,
it would be the maximum likelihood (ML) estimator of $\eta$,
it would again be unbiased, and its MSE would be $\eta/\lambda$.
The factor of $(1+\eta)$ excess seen in \eqref{eq:MSE-baseline}
is the cost of the randomness of a Poisson ion beam.
We will see approximately this factor of improvement from time-resolved measurement,
thus approximately cancelling the effect of source shot noise.

\subsection{Oracle estimator}
\label{ssec:oracle}
If one were able to know $M$, the estimate
\begin{equation}\label{eq:eta_hat_oracle}
  \etaOracle(Y,M) = \frac{Y}{M}
\end{equation}
would be superior to $\etaBaseline$
because $Y$ is the sum of $M$ random variables,
each with mean $\eta$.
One can view $\etaOracle$ as mitigating the source shot noise
by using the exact number of ions.
Along with the issue of resolving $0/0$ when no ions are incident,
the problem with this is that $M$ is not observable.
While the exact number of ions $M$ cannot be known exactly from only observing $Y$,
we will see that $M$ becomes approximately known with
time-resolved measurement.

For a non-Bayesian analysis of $\etaOracle$,
we can fix an arbitrary value $\eta_0$ as the estimate produced when $M = 0$.
While $\etaOracle$ is unbiased whenever $M > 0$
(which can be seen by iterated expectation with conditioning on $M$),
there is nothing computable from the data $(Y, M) = (0, 0)$
that makes $\etaOracle$ unbiased overall.
Specifically,
\begin{align}
\bias(\etaOracle)
&= \iE{\etaOracle(Y,M)} - \eta \nonumber \\
&\eqlabel{a} \iE{ \iE{\etaOracle(Y,M) \smid M} } - \eta \nonumber \\
&\eqlabel{b} \eta_0 \, \P{M = 0} + \eta (1 - \P{M = 0}) - \eta \nonumber \\
&\eqlabel{c} \eta_0 e^{-\lambda} + \eta(1 - e^{-\lambda}) - \eta \nonumber \\
&= (\eta_0 - \eta)e^{-\lambda},
\label{equ:oracle_bias}
\end{align}
where (a) follows from the law of iterated expectation;
(b) from $\iE{ \etaOracle(Y,M) \smid M = m }$ taking only the values $\eta_0$ and $\eta$; and
(c) from the Poisson distribution of $M$.
The variance of the estimate is
\begin{align}
\var{\etaOracle}
&\eqlabel{a} \E{\var{\etaOracle(Y,M) \smid M}}
           + \var{\E{\etaOracle(Y,M) \smid M}} \nonumber \\
&\eqlabel{b} \eta \sum_{m = 1}^{\infty} \frac{1}{m}e^{-\lambda}\frac{\lambda^m}{m!}
           + e^{-\lambda}(1 - e^{-\lambda})(\eta - \eta_0)^2 \nonumber \\
&\eqlabel{c} \eta g(\lambda) + e^{-\lambda}(1 - e^{-\lambda})(\eta - \eta_0)^2,
\label{equ:oracle_var}
\end{align}
where (a) follows from the law of total variance;
(b) from the conditional distribution of $\etaOracle$ being the constant $\eta_0$ for $M =0$ and the sample mean of $m \, \text{Poisson}(\eta)$ random variables for $M = m$, $m >0$; and
(c) introduces a function
$g(\lambda) = \sum_{m = 1}^{\infty}(\Frac{1}{m}) e^{-\lambda}\Frac{\lambda^m}{m!}$, which has no elementary closed form.
Notice that $g(\lambda) \approx \lambda$ for $\lambda \ll 1$, since only the $m = 1$
term is appreciable; moreover, it can be shown that $g(\lambda) \approx 1/\lambda$ for $\lambda \gg 1$.

The bias and variance computations can be combined to give an expression for the MSE of the oracle estimator:
\begin{align}
\mathrm{MSE}(\etaOracle)
&= [\bias(\etaOracle)]^2 + \var{\etaOracle} \nonumber \\
&\eqlabel{a} \left[ (\eta_0 - \eta)e^{-\lambda}\right]^2 + \eta g(\lambda) + e^{-\lambda}(1 - e^{-\lambda})(\eta - \eta_0)^2 \nonumber \\
&= \eta g(\lambda) + e^{-\lambda} (\eta - \eta_0)^2,
\label{equ:MSE_oracle}
\end{align}
where (a) follows by substituting~\eqref{equ:oracle_bias} and~\eqref{equ:oracle_var}. Furthermore,
\begin{equation}
\mathrm{MSE}(\hat{\eta}_{\rm{oracle}}) \geq \eta g(\lambda), 
\end{equation}
with the bound achieved when $\eta_0 = \eta$. We stress that this bound is unachievable because $\eta$ is not \textit{a priori} known.

\subsection{Fisher information}
\label{ssec:FI-analysis}
The MSE of any unbiased estimator is lower bounded by the reciprocal
of the Fisher information via the {\CR} bound (CRB)~\cite{Kay1993}.
Fisher information is also central to our explanation of why
time-resolved measurement combined with ML estimation greatly mitigates
source shot noise.

The Fisher information for the estimation of $\eta$ from $Y$ in the Poisson--Poisson model,
with $\lambda$ a known parameter, can be simplified to
\begin{align} 
\mathcal{I}_Y(\eta \sMid \lambda)
&= \E{ \left(\frac{\partial \log P_Y(Y \sMid \eta, \lambda)}{\partial \eta} \right)^2 \sMid \eta } \nonumber \\
&= \sum_{y = 0}^{\infty} \left(
     \frac{y}{\eta} - \frac{P_Y(y + 1 \sMid \eta, \lambda)}{P_Y(y \sMid \eta, \lambda)} \frac{y+1}{\eta}
     \right)^2 P_Y(y \sMid \eta, \lambda);
\label{equ:Fisher_info_equation_PP}
\end{align}
see \ref{app:FI_PP_expression} for a derivation.
While this expression is not readily comprehensible,
it can be used to compute $\mathcal{I}_Y(\eta \sMid \lambda)$ numerically and
to derive certain useful asymptotic approximations and limits.

\begin{figure}
  \begin{center}
    \includegraphics[width=0.9\linewidth]{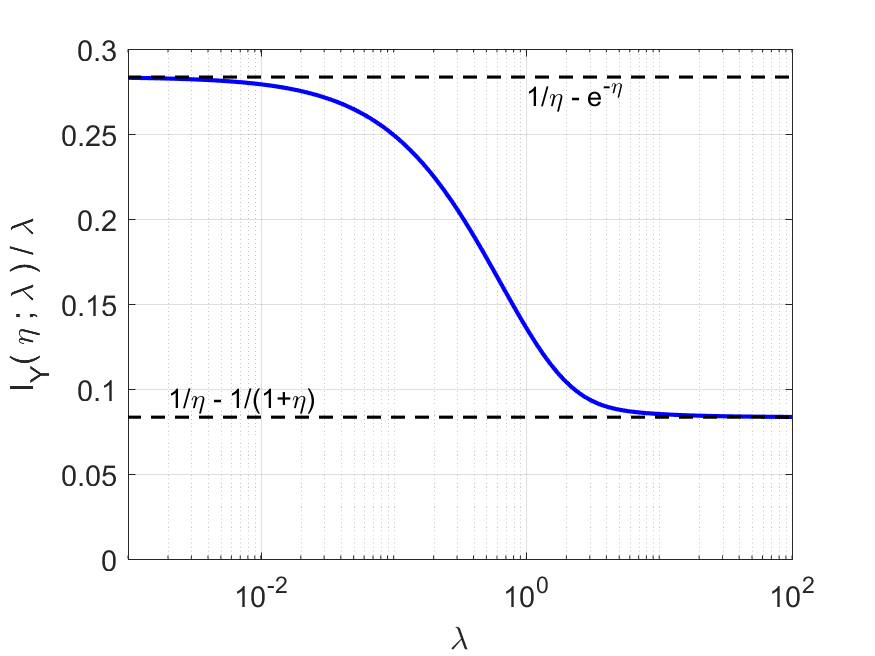}
  \end{center}
  \caption{Normalized Fisher information $\mathcal{I}_Y(\eta \sMid \lambda)/\lambda$
        as a function of $\lambda$ for $\eta = 3$.
        Low-dose measurements are more informative per incident ion than high-dose measurements.
        The marked asymptotes are derived in \ref{app:FI_Neyman_limits}.}
  \label{fig:normalized_Fisher_info}
\end{figure}

One can study $\mathcal{I}_Y(\eta \sMid \lambda)/\lambda$ as the
information gain per incident ion.
As illustrated in Figure~\ref{fig:normalized_Fisher_info},
this \emph{normalized Fisher information} (NFI) 
is a decreasing function of $\lambda$,
with
\begin{equation}\label{eq:FI-low-lambda}
  \lim_{\lambda \rightarrow 0}
    \frac{\mathcal{I}_Y(\eta \sMid \lambda)}{\lambda}
    = \frac{1}{\eta} - e^{-\eta}
\end{equation}
and
\begin{equation}\label{eq:FI-high-lambda}
  \lim_{\lambda \rightarrow \infty}
    \frac{\mathcal{I}_Y(\eta \sMid \lambda)}{\lambda}
    = \frac{1}{\eta(1+\eta)}
    = \frac{1}{\eta} - \frac{1}{1+\eta},
\end{equation}
as derived in \ref{app:FI_Neyman_limits}.
The ratio of these limits is
\begin{equation}\label{eq:FI-gain-factor}
  \beta(\eta) = (1+\eta)( 1 - \eta e^{-\eta} ),
\end{equation}
which varies from $1$ to $\approx 1+\eta$ as $\eta$ increases from $0$.
Recall the $1+\eta$ factor arose
in Section~\ref{ssec:baseline}
as the cost of randomness of a Poisson ion beam.

Comparing \eqref{eq:FI-high-lambda} with \eqref{eq:MSE-baseline}, we see that,
asymptotically for large $\lambda$,
the baseline estimator achieves the CRB\@.
In contrast,
for low $\lambda$,
the probability for $M=0$ is appreciable,
so there is no (even approximately) unbiased estimator.

\section{Time-resolved measurements}
\label{sec:TR}

Taken together, the analyses in Section~\ref{sec:Model_methods} suggest that
there may be a way for the baseline estimate from \eqref{eq:eta-baseline}
to be improved upon to give a reduction in MSE by the factor in \eqref{eq:FI-gain-factor}.
Time-resolved measurement indeed achieves this improvement.
We examine this first through Fisher information and then
through simulated performance of the ML estimator for imaging.

\subsection{Fisher information ratios}
\begin{figure}
  \begin{center}
    \includegraphics[width=0.9\linewidth]{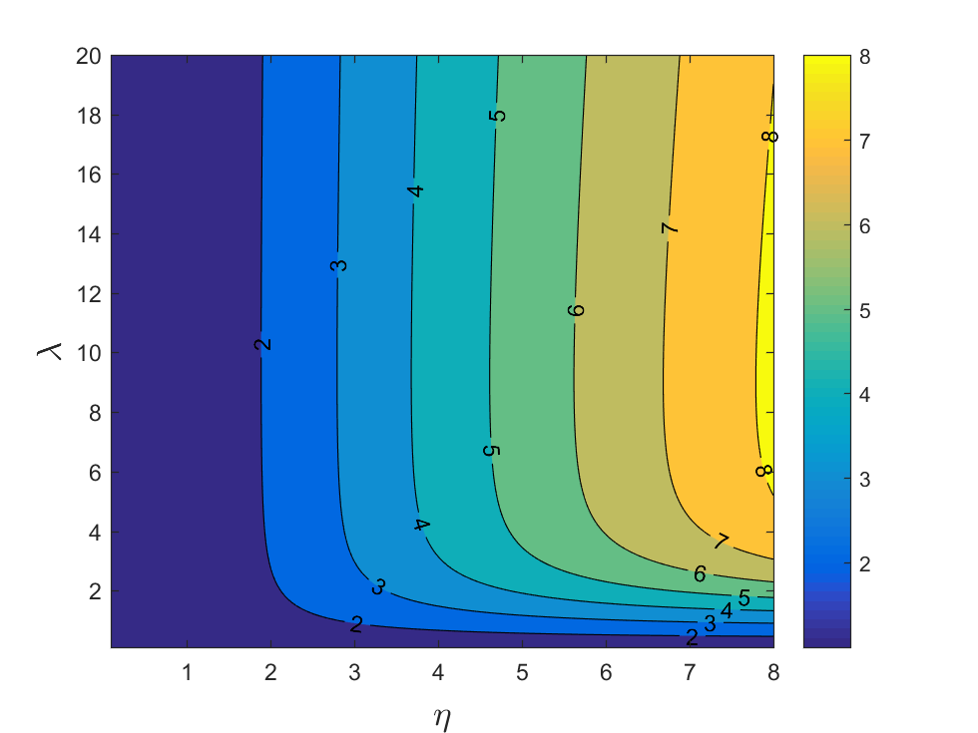}
  \end{center}
  \caption{Fisher information ratio for Poisson--Poisson model used to predict the
	multiplicative factor by which MSE would be improved with
	the introduction of time-resolved measurement into FIB microscopy ($n = 500$).}
  \label{fig:Fisher_info_PP}
\end{figure}

If we divide pixel dwell time $t$ into $n$ sub-acquisitions to obtain
$Y_1,\,Y_2,\,\ldots,\,Y_n$, these are independent and obtained with $\lambda$
replaced by $\lambda/n$.
The FI for the set of sub-acquisitions together is
\begin{align}
\mathcal{I}^{\rm TR}_Y (\eta \sMid \lambda, n)
 &\eqlabel{a} n \, \mathcal{I}_Y (\eta \sMid \lambda/n)
   \label{eq:Fisher_info_TR} \\
 &= \lambda \, \frac{\mathcal{I}_Y (\eta \sMid \lambda/n)}{\lambda/n} \nonumber \\
 &\approxlabel{b} \lambda\left( \frac{1}{\eta} - e^{-\eta} \right),
\label{equ:Fisher_info_approx_TR}
\end{align}
where (a) follows from the additivity of FI over independent observations; and
(b) holds for large enough $n$ because of \eqref{eq:FI-low-lambda}.
Without time-resolved measurement,
for total dose values useful for imaging (say, $\lambda > 2$),
as illustrated in Figure~\ref{fig:normalized_Fisher_info},
the limit in \eqref{eq:FI-high-lambda} provides a good approximation of the FI:
\begin{equation}
\mathcal{I}_Y(\eta \sMid \lambda)
  \approx \lambda \left( \frac{1}{\eta} - \frac{1}{1+\eta} \right).
\label{equ:Fisher_info_approx}
\end{equation}
The ratio of \eqref{equ:Fisher_info_approx_TR} and \eqref{equ:Fisher_info_approx}
was already computed as \eqref{eq:FI-gain-factor}.
This ratio
gives a convenient way to evaluate the improvement from time-resolved measurement and data processing.
Figure~\ref{fig:Fisher_info_PP} is a contour plot of the FI ratio
(without using approximations)
for
$n = 500$.
The ratio of Fisher informations is the reciprocal of the ratio of {\CR} lower bounds.
For example, where the contour is labeled 5,
splitting the fixed dose $\lambda$ into $n = 500$ sub-acquisitions enables the
reduction of the mean-squared errors (MSE) to 20\% of the MSE value from a single full-dose experiment.

\subsection{{\CR} bounds}
\label{ssec:CRB_numerical}
\begin{figure}
\begin{subfigure}{.5\linewidth}
  \centering
  \includegraphics[width=\linewidth]{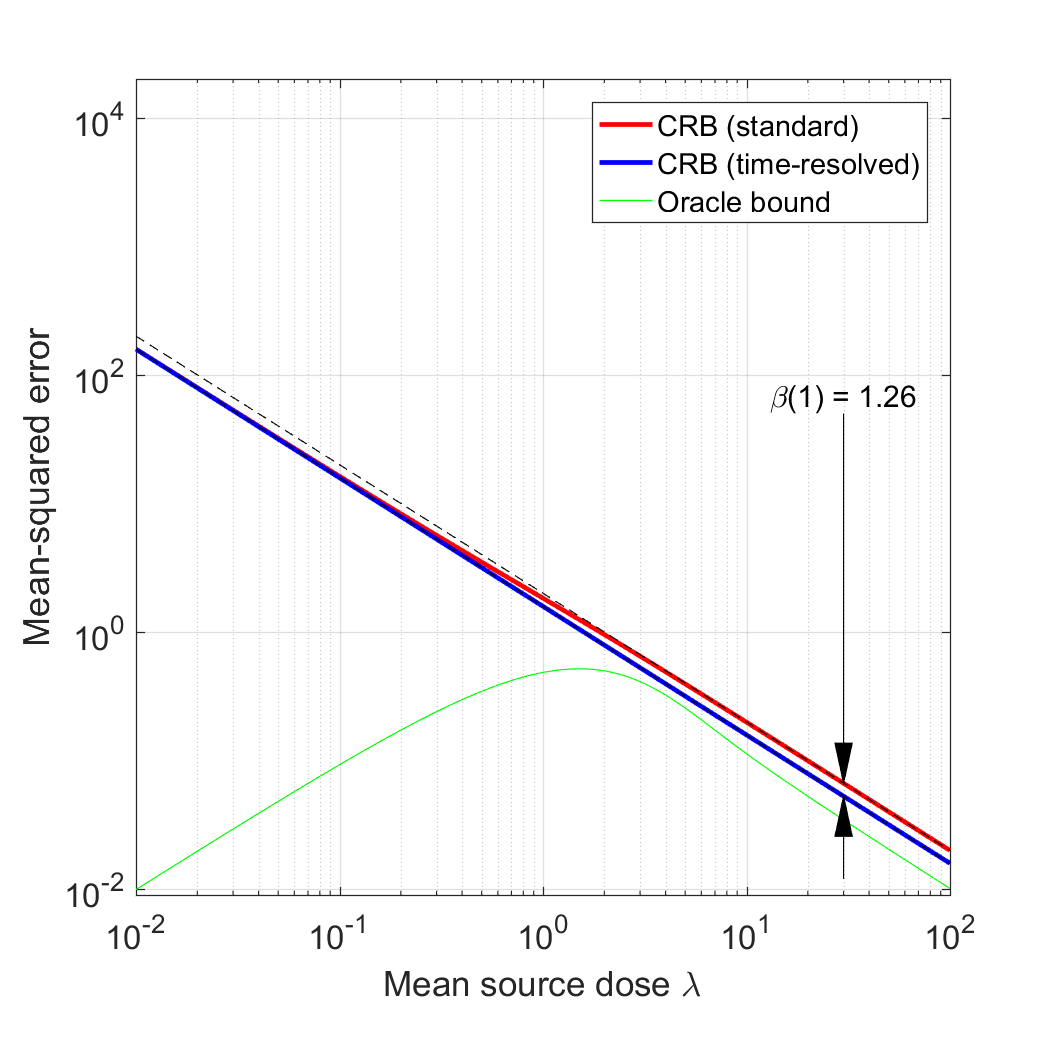}
  \caption{$\eta = 1$}
  \label{fig:beta_eta_1}
\end{subfigure}%
\begin{subfigure}{.5\linewidth}
  \centering
  \includegraphics[width=\linewidth]{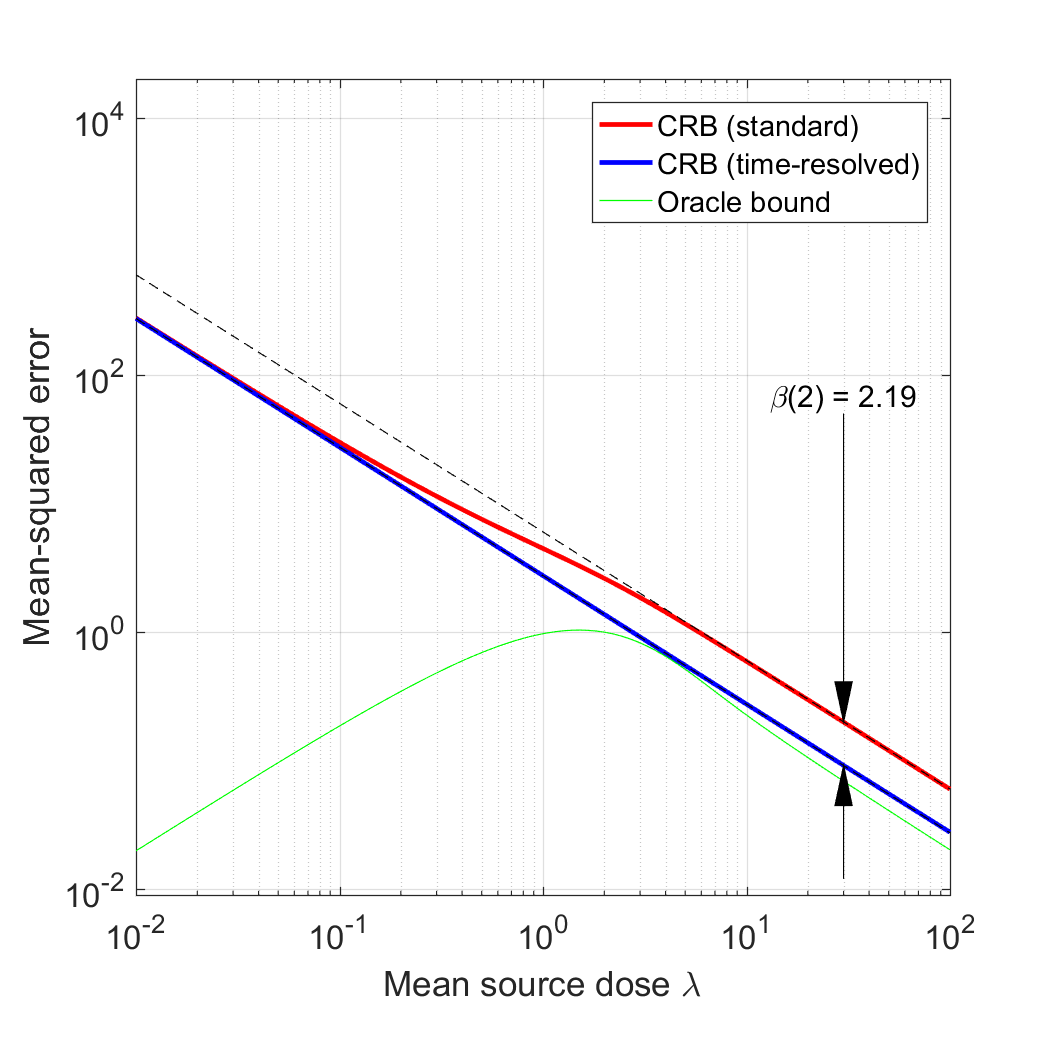}
  \caption{$\eta = 2$}
  \label{fig:beta_eta_2}
\end{subfigure}
\begin{subfigure}{.5\linewidth}
  \centering
  \includegraphics[width=\linewidth]{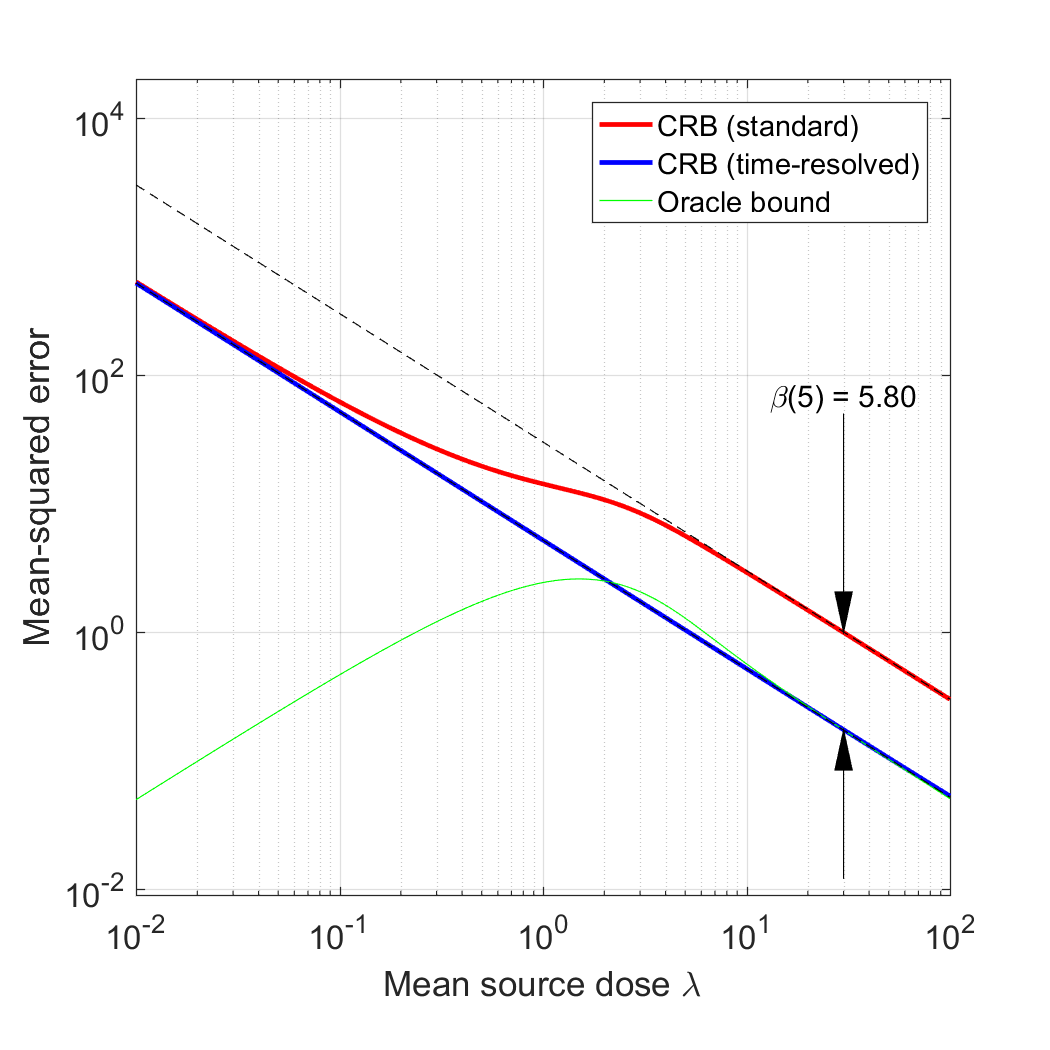}
  \caption{$\eta = 5$}
  \label{fig:beta_eta_5}
\end{subfigure}%
\begin{subfigure}{.5\linewidth}
  \centering
  \includegraphics[width=\linewidth]{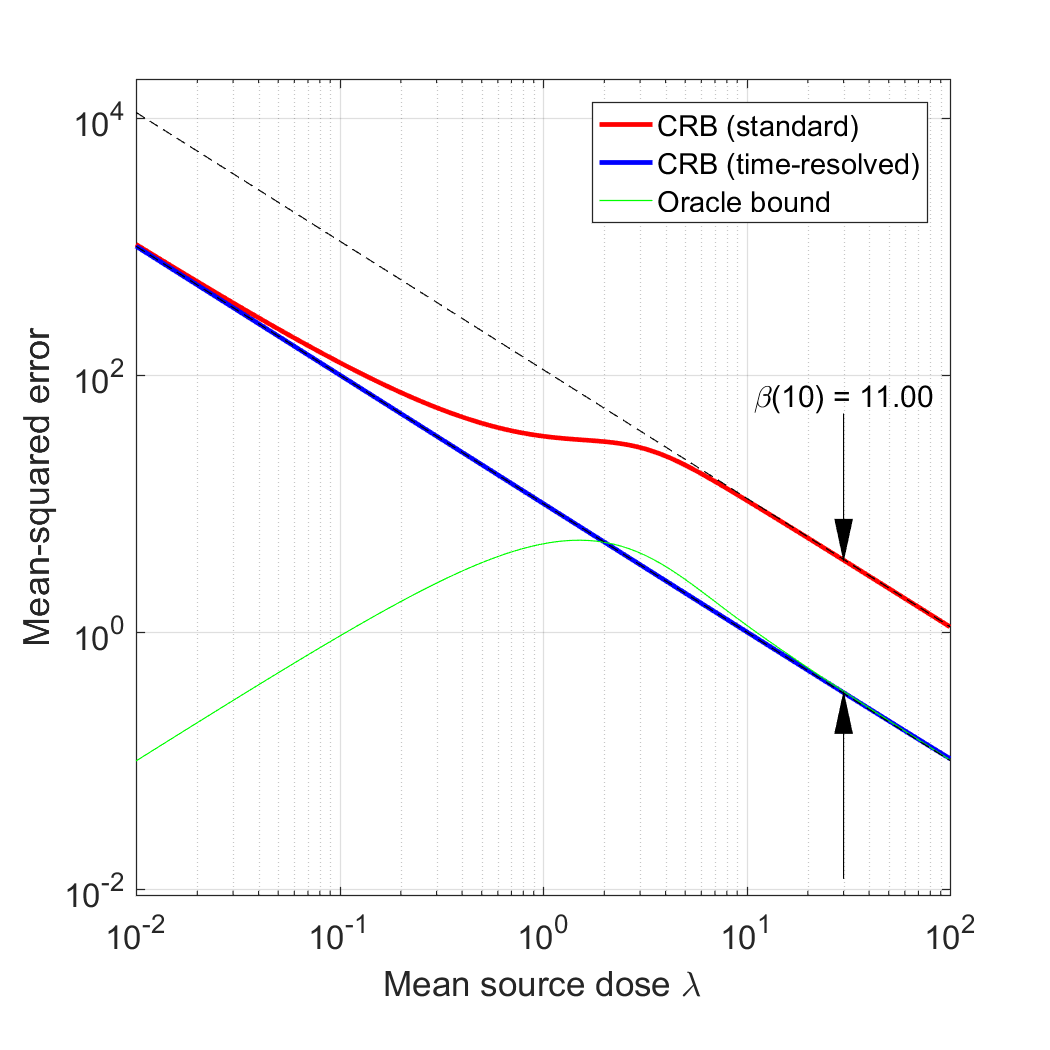}
  \caption{$\eta = 10$}
  \label{fig:beta_eta_10}
\end{subfigure}
\caption{Comparison between the {\CR} bounds obtained for conventional method (red, see (\ref{equ:Fisher_info_equation_PP})) and time-resolved measurement (blue, $n=10\,000$, see (\ref{eq:Fisher_info_TR})) for several values of mean secondary electron yield $\eta$.
Also shown in each plot is the oracle bound \eqref{equ:MSE_oracle} for the estimator \eqref{eq:eta_hat_oracle};
recall that this bound is based on estimator that is "increasingly unimplementable" as $\lambda \rightarrow 0$ since it is derived from assuming $\hat{\eta} = \eta$ when no ions are incident.
Each plot also shows the performance from (\ref{eq:MSE-baseline}) for the baseline estimate, which is a high-$\lambda$ asymptote for the {\CR} bound in the case of conventional sensing.
The expression (\ref{equ:MSE_TR}) is plotted as well, but it lies coincident with the blue curve.}
\label{fig:beta_eta_comparison}
\end{figure}

The CRB informs us that no unbiased estimator can have variance lower than the reciprocal of the Fisher information.
Thus, the FI for a single measurement \eqref{equ:Fisher_info_equation_PP}
and for time-resolved measurement \eqref{eq:Fisher_info_TR}
imply bounds on MSE for unbiased estimators,
as plotted in Figure~\ref{fig:beta_eta_comparison}.
The asymptotic approximation \eqref{equ:Fisher_info_approx_TR}
implies a bound that applies to any unbiased estimator $\etaTR$
computed from the time-resolved measurements:
\begin{equation}
\mathrm{MSE}(\etaTR) \geq \frac{\eta/(1 - \eta e^{-\eta})}{\lambda}.
\label{equ:MSE_TR}
\end{equation}
For the performance without time-resolved measurement,
this should be contrasted with \eqref{eq:MSE-baseline};
the baseline estimator achieves the CRB asymptotically in large $\lambda$,
and Figure~\ref{fig:normalized_Fisher_info}
illustrates that the asymptote is a good approximation for
values of $\lambda$ useful for imaging.

\subsection{Joint distribution and ML estimation}
For time-resolved measurements, the joint distribution is
\begin{equation}
P_{Y_1,\, \ldots,\, Y_n}(y_1,\, \ldots,\, y_n \sMid \eta, \lambda)
= \prod_{k=1}^{n}
P_{Y}(y_k \sMid \eta, \lambda/n ),
\label{eq:TRsensing-PMF}
\end{equation}
where $P_Y(\cdot \sMid \cdot, \cdot)$ is given by \eqref{equ:neyman}.
Roughly speaking, when the sub-acquisitions are short enough (that is, $n$ is large enough), 
each
sub-acquisition will have very low dose and thus very likely have
0 or 1
incident ion.
Assuming most sub-acquisitions with 1 incident ion yield at least 1 SE,
one can use the number of sub-acquisitions with a strictly positive number
of detected SEs as a proxy for the number of ions $M$.
This gives some plausibility for mitigating source shot noise
and is the intuitive justification of the ``quotient mode'' developed by Zeiss~\cite{Notte2013}.
Our
methods use the more precise model \eqref{eq:TRsensing-PMF}.
Most importantly, we account for the probability of an 
incident ion resulting in zero detected SEs.

Given the observation $(y_1,\,y_2,\,\ldots,\,y_n)$,
the ML estimate for $\eta$ is
\begin{equation}
\hat{\eta}_{\rm TR} = \argmax_{\eta} \, P_{Y_1,  \,\ldots, \, Y_n}(y_1, \,\ldots, \, y_n \sMid \eta, \lambda).
\label{eq:eta_hat_TR_PP}
\end{equation}
Since 
$P_{Y_1,\, \ldots,\, Y_n}(y_1,\, \ldots,\, y_n \sMid \eta, \lambda)$
is a non-convex function of $\eta$,
we compute the optimization via grid search.
This is not prohibitively complex because the decision variable is scalar.

\subsection{Synthetic numerical results}
Simulation results
also demonstrate the improvement gained from 
time-resolved data acquisition and processing.
For a fixed dose, a lower
reconstruction MSE compared to the conventional method is obtainable;
equivalently, time-resolved measurement gives similar imaging MSE
with a reduced ion dose compared to the conventional method.

\begin{figure}
 \centering
  \begin{tabular}{cc}
    \begin{subfigure}{0.4\linewidth}
      \includegraphics[width=\linewidth]{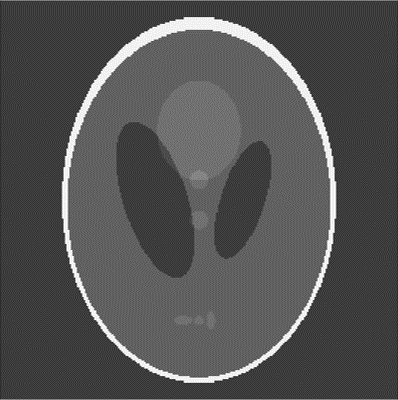}
      \captionsetup{justification=centering}
      \caption{\small \footnotesize ground truth \\ ~}
      \label{fig:shepp_logan_PP_groundtruth}
    \end{subfigure} &
    \begin{subfigure}{0.4\linewidth}
      \includegraphics[width=\linewidth]{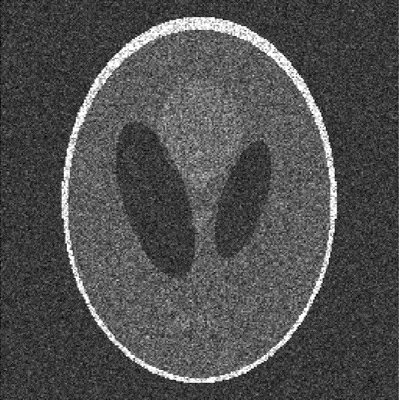}
      \captionsetup{justification=centering}
      \caption{\footnotesize conventional \\ \footnotesize $\lambda = 20$, MSE: 0.5934}
      \label{fig:shepp_logan_PP_conventional}
    \end{subfigure} \\
    \begin{subfigure}{0.4\linewidth}
      \includegraphics[width=\linewidth]{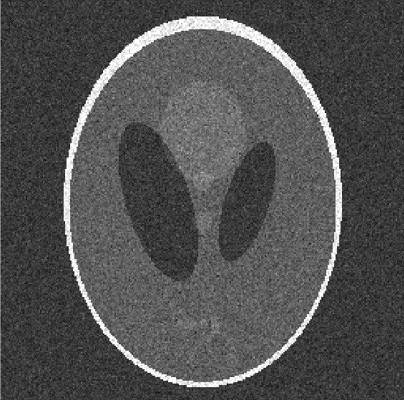}
      \captionsetup{justification=centering}
      \caption{\footnotesize time-resolved \\ \footnotesize $\lambda = 20$, MSE: 0.2297}
      \label{fig:shepp_logan_PP_TR}
    \end{subfigure} &
    \begin{subfigure}{0.4\linewidth}
      \includegraphics[width=\linewidth]{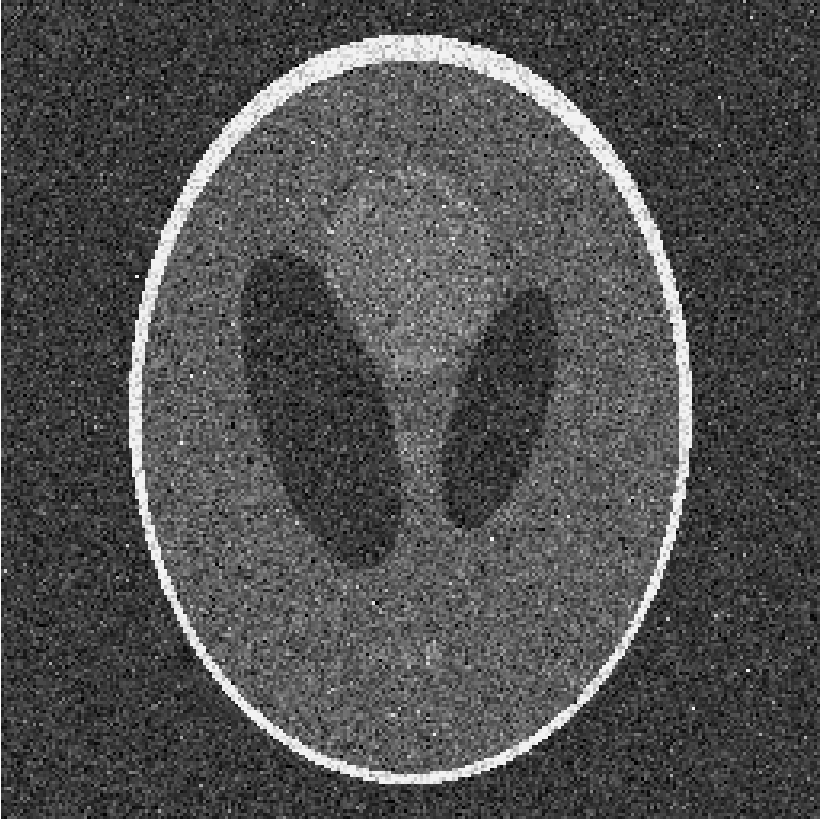}
      \captionsetup{justification=centering}
      \caption{\footnotesize time-resolved \\ \footnotesize $\lambda = 10$, MSE: 0.513}
      \label{fig:dose_reduc_shepp_logan_PP_TR}
    \end{subfigure}
  \end{tabular}
  \\[-2mm]
 \caption{Simulated HIM experiment for a sample with mean secondary electron yield in $[2,\,8]$ for Poisson--Poisson (direct electron detection) model in Section~\ref{subsec:PPModel}.
 (a) Ground truth image.
 (b) Conventional HIM image with $\lambda = 20$.
 (c) and (d) Pixelwise ML estimates \eqref{eq:eta_hat_TR_PP}
 computed from $n=100$ time-resolved measurements
 with $\lambda = 20$ and $\lambda = 10$.
 Comparing (b) and (c) demonstrates MSE reduction at fixed dose.
 Comparing (b) and (d) demonstrates dose reduction without increase of MSE\@.
 These results do not use spatial regularization.}
 \label{fig:shep_logan_PP}
\end{figure}

Figure~\ref{fig:shepp_logan_PP_groundtruth}
shows the ``'Modified Shepp--Logan phantom''
provided by the Matlab {\tt phantom} command,
at size $256 \times 256$,
scaled to give ground truth SE values in the interval $[2,\,8]$,
as suggested in~\cite{Notte2006_HIM}.
Figures~\ref{fig:shepp_logan_PP_conventional}
and~\ref{fig:shepp_logan_PP_TR}
show that for a fixed dose of $\lambda = 20$,
time-resolved measurement with $n=100$ sub-acquisitions achieves
an MSE reduction by a factor of 2.4\@.

An alternative way
to demonstrate the improvement due to time-resolved measurement
is through
a dose reduction for fixed image quality.
The proposed time-resolved measurement reconstruction,
shown in Figure~\ref{fig:dose_reduc_shepp_logan_PP_TR},
achieves a slightly lower MSE
than the conventional reconstruction in Figure~\ref{fig:shepp_logan_PP_conventional}
with a dose of only 10~ions per pixel.

\section{Hierarchical compound models}
\label{sec:hierarchical_models}
The model introduced in Section~\ref{subsec:PPModel} assumes direct secondary electron counting, so that the number of SEs is the final readout of the device. 
In current HIM instruments, the output is more indirect.
We now discuss some plausible models for the SE detection process and show
that FI-based analysis continues to suggest substantial advantages for time-resolved measurement.

\subsection{Poisson--Poisson--Normal}
\label{subsec:PPNModel}
In a typical HIM instrument, SEs emitted due to ion-sample interaction are accelerated towards a phosphor scintillator plate by an electric field.
Photons generated as a result of SE-scintillator interaction are amplified by a photomultiplier tube (PMT) and subsequently converted into an electrical current \cite{notte2007introduction}.
There is high degree of randomness in the scintillator and the PMT response~\cite{hakamataphotomultiplier}, both of which cause randomness in the output current.

As one possible model with only two additional parameters,
one could model the contribution to the final measurement from each detected SE as being normally distributed.
Specifically, suppose the measured output current due
to the $j$th SE is normal with mean $c_1$ and variance $c_2$, i.e., $Z_j \sim \mathcal{N}(c_1, c_2)$.
Then, the observation model at one pixel becomes
\begin{equation}
U = \sum_{j = 1}^{Y} Z_j,
\end{equation}
where $Y$ is the number of SEs.
Combining the normal distribution with the Neyman Type A distribution in \eqref{equ:neyman}
gives the following probability density function (PDF) for $U$:
\begin{equation}
f_U(u \sMid \eta, \lambda, c_1, c_2) 
= \sum_{y = 1}^{\infty} \frac{1}{\sqrt{2 \pi c_2 y}}
   \exp\!\left(-\frac{(u - c_1y) ^ 2}{2 c_2 y}\right)
   P_{Y}(y \sMid \eta, \lambda).
\label{eq:PDF_PPNdistr}
\end{equation}

Under \eqref{eq:PDF_PPNdistr}, the ML estimate of $\eta$, from $n$ short acquisitions, becomes:
\begin{equation}
	\hat{\eta}_{\rm TR} = \argmax_{\eta} \, f_{U_1,  \,\ldots, \, U_n}(u_1, \,\ldots, \, u_n \sMid \eta, \lambda, c_1, c_2),
\label{eq:eta_hat_TR_PPN}
\end{equation}
where 
\[
f_{U_1,\ldots,U_n}(u_1, \,\ldots, \, u_n \sMid \eta, \lambda, c_1, c_2)
= \prod_{k = 1}^{n} f_U(u_k \sMid \eta, \lambda, c_1, c_2).
\]

\subsection{Quantized Poisson--Poisson--Normal}
\label{subsec:PPN_quantized_Model}
While the Poisson--Poisson--Normal model
of Section~\ref{subsec:PPNModel}
attempts to account for randomness in the scintillator and PMT responses,
several aspects of a typical HIM instrument are not modelled.
In particular,
\eqref{eq:PDF_PPNdistr} allows negative measurements and
the analog-to-digital conversion (ADC) to map output current into an 8-bit gray scale value is unmodelled.
Assuming analog gains are set to avoid ADC overload,
both of these effects can be accounted for by rounding the measurement to its nearest nonnegative integer.
(Overload could be accounted for similarly.)
Consequently, the PMF for the observed output $\Utilde \in \mathbb{N}$ for each pixel is then:
\begin{equation}
P_{\Utilde}(\utilde \sMid \eta, \lambda, c_1, c_2) 
= \dfrac{\int_{\utilde - \frac{1}{2}}^{\utilde + \frac{1}{2}} f_U(u \sMid \eta, \lambda, c_1, c_2) \, \mathrm{d}u}
        {\int_{-\frac{1}{2}}^{\infty} f_U(u \sMid \eta, \lambda, c_1, c_2) \, \mathrm{d}u}.
\label{eqn:PPNQ_PMF}
\end{equation}
Note that the denominator in \eqref{eqn:PPNQ_PMF}
normalizes the PMF to account for there being no negative measurements.
The corresponding ML estimate $\hat{\eta}_{\rm TR}$ under this new model can be written in a similar fashion to \eqref{eq:eta_hat_TR_PPN}.

\begin{figure}
  \begin{center}
    \includegraphics[width=0.9\linewidth]{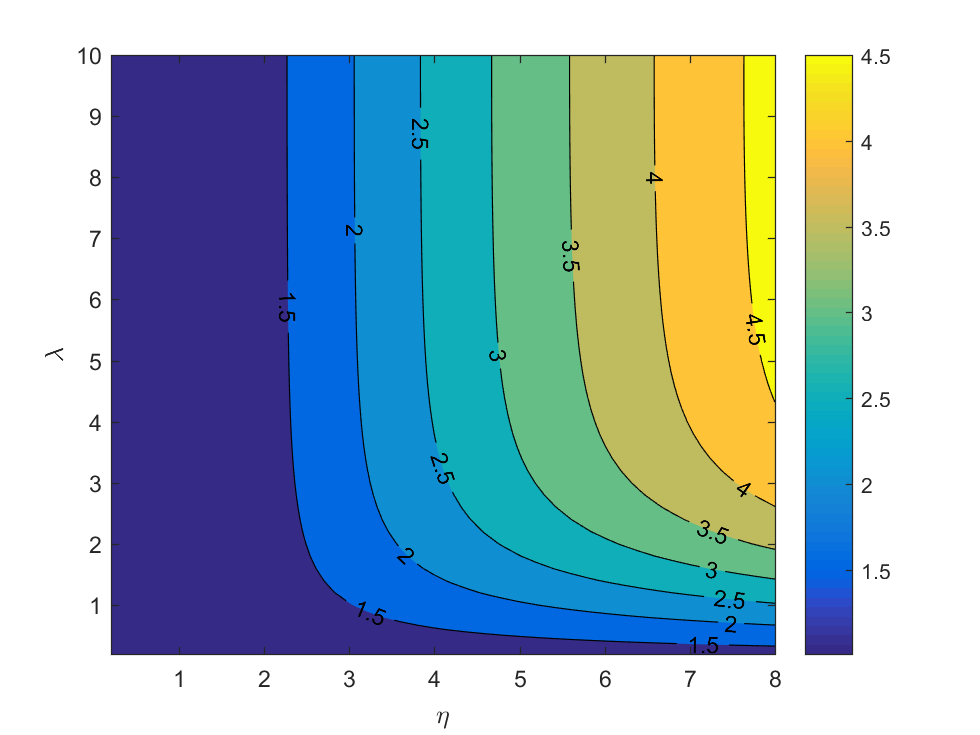}
  \end{center}
  \caption{Fisher information ratio for Quantized Poisson--Poisson--Normal model
    used to predict the multiplicative factor by which MSE would be improved with the introduction of time-resolved measurement into FIB microscopy
    ($c_1 = 10$, $c_2 = 100$, and $n = 500$).}
  \label{fig:Fisher_info_PPNQ}
\end{figure}

The FI for the estimation of $\eta$ from $\Utilde$,
with $\lambda$, $c_1$, and $c_2$ as known parameters, is
\begin{equation} 
\mathcal{I}_\Utilde(\eta \sMid \lambda, c_1, c_2)
=\E{ \left(\frac{\partial \log P_\Utilde(\utilde \sMid \eta, \lambda, c_1, c_2)}{\partial \eta} \right)^2 \sMid \eta } .
\label{equ:Fisher_info_equation_QPPN}
\end{equation}
Though we have no insightful simplifications or approximations of
$\mathcal{I}_\Utilde(\eta \sMid \lambda, c_1, c_2)$,
we can compare it numerically to
$n\,\mathcal{I}_\Utilde(\eta \sMid \lambda/n, c_1, c_2)$
to quantify the increase in information from TR measurements.
A contour plot showing the Fisher information ratio for TR versus conventional data acquisition is given in
Figure~\ref{fig:Fisher_info_PPNQ} under this new Quantized Poisson--Poisson--Normal hierarchical model.
The plot suggests that improvement MSE improvements are still obtainable by using TR data.
However, comparing it with Figure~\ref{fig:Fisher_info_PP},
it is clear that the overall possible gain is reduced in this new model.
This is attributable to the extra layer of randomness introduced by the scintillator and PMT\@.
In addition, this discrepancy can be viewed as theoretical support for preferring direct secondary electron counting, over other methods of electron detection.

\begin{figure}
 \centering
  \begin{tabular}{cc}
   \begin{subfigure}{0.4\linewidth}
   \centering
    \includegraphics[width=\linewidth]{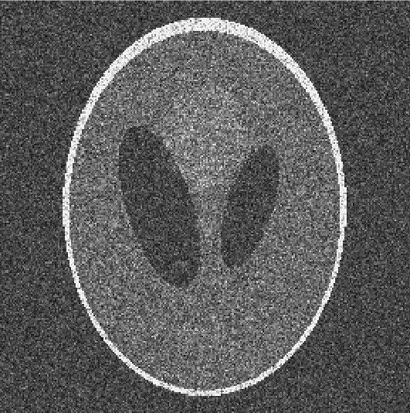}
    \captionsetup{justification=centering}
    \caption{\footnotesize conventional \\ \footnotesize $\lambda = 20$, MSE: 1.053}
    \label{fig:shepp_logan_PPN_conventional_lambda_20}
   \end{subfigure} &
   \begin{subfigure}{0.4\linewidth}
   \centering
    \includegraphics[width=\linewidth]{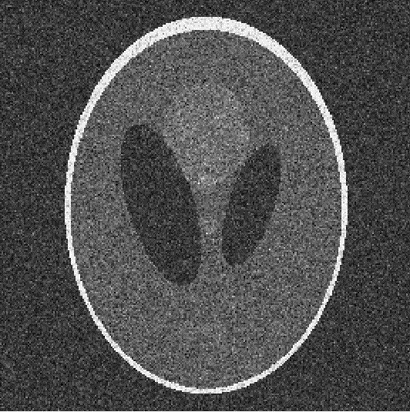}
    \captionsetup{justification=centering}
    \caption{\footnotesize time-resolved \\ \footnotesize $\lambda = 20$, MSE: 0.562}
    \label{fig:shepp_logan_PPN_TR_lambda_20}
   \end{subfigure}
  \end{tabular}
 \\[-2mm]
 \caption{Simulated HIM experiment for
  Quantized Poisson--Poisson--Normal model in Section~\ref{subsec:PPN_quantized_Model} when
  $\lambda = 20$, $c_1 = 10$, and $c_2 = 200$:
  (a) Conventional HIM image.
  (b) Pixelwise ML estimates computed from $n = 100$ time-resolved measurements.
  These results do not use spatial regularization.}
 \label{fig:shep_logan_PPNQ_lambda_20}
\end{figure}

Figure~\ref{fig:shep_logan_PPNQ_lambda_20} shows the results of simulations
for the same sample as in Figure~\ref{fig:shep_logan_PP}.
At the same dose of $\lambda = 20$, the MSEs are higher than in
Figure~\ref{fig:shep_logan_PP}, but substantial improvement from
time-resolved measurement is again demonstrated.

\subsection{Poisson--Poisson--Poisson}
One final model further illustrates the flexibility of hierarchical modeling and the general potential of TR measurement.
Removing the use of a normal distribution to model phosphor and PMT response,
suppose that photons emitted by the scintillator can be directly measured
instead of being converted into an electrical signal.
Through the use of a time-resolved single-photon detector, we can count the number of emitted photons;
for instance, a single photon avalanche diode (SPAD) detector with time-correlated single photon counting could be used.
Modeling the number of photons generated due to the $j$th SE
as a Poisson random variable $W_j$ with mean $c$
and the  observation at one pixel by
\begin{equation}
V = \sum_{j = 1}^{Y} W_j,
\end{equation}
the PMF for the final read-out becomes
\begin{equation}
P_V(v \sMid \eta, \lambda, c) = \sum\limits_{y = 0}^{\infty} \cfrac{e^{-cy} (cy)^v}{v!} \frac{e^{-\lambda} \eta^y}{y!} \sum\limits_{m = 0}^{\infty} \frac{(\lambda e^{-\eta}) ^m m^y}{m!}.
\label{eq:PPP-PMF}
\end{equation}
Equation~\eqref{eq:PPP-PMF} is be obtained
analogously to \eqref{eq:PDF_PPNdistr}
by combining the Poisson distribution of $W_j$ with the
Neyman Type A distribution of $Y$ given in \eqref{equ:neyman}.

\begin{figure}
\centering
\includegraphics[width=0.9\linewidth]{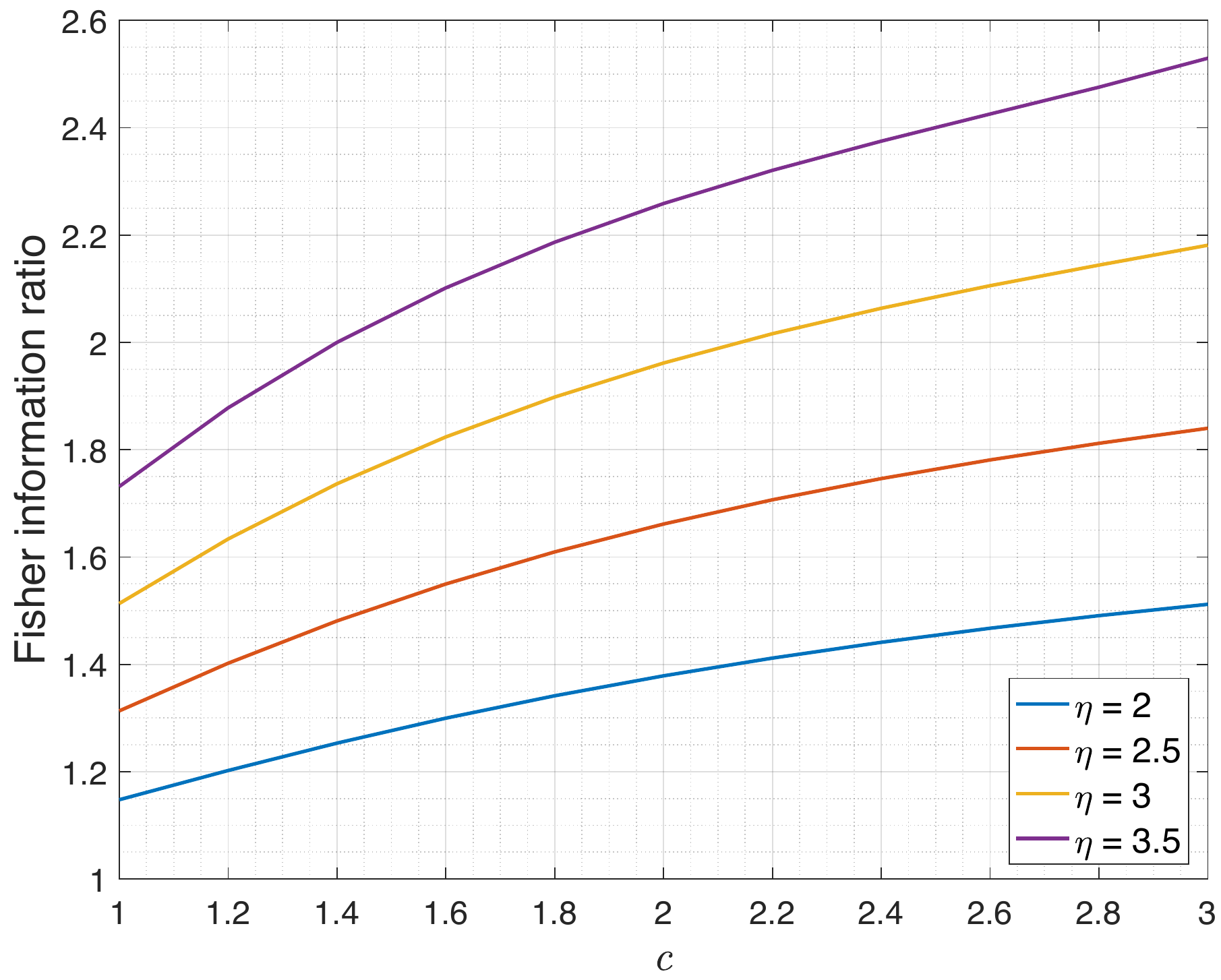}
  \caption{Fisher information ratio for Poisson--Poisson--Poisson model
    used to predict the multiplicative factor by which MSE would be improved with the introduction of time-resolved measurement into FIB microscopy
    (four values of $\eta$, $n = 500$, $\lambda = 2$).}
  \label{fig:Fisher_info_PPP}
\end{figure}

Figure~\ref{fig:Fisher_info_PPP} shows plots of the Fisher information ratio for TR measurements
under this Poisson--Poisson--Poisson model
as a function of $c$ for four $\eta$ values when $\lambda = 2$.
The plots show that improvements in MSE for a fixed dose
(or dose reduction for a desired MSE)
is expected when the proposed time-resolved sensing method is used.

\section{HIM imaging results}
\label{sec:HIM-results}
\subsection{Experiment details}
\begin{figure*}
  \centering
  \begin{tabular}{c@{\,\,}c@{\,\,}c@{\,\,}c@{\,\,}c}
    \begin{subfigure}{0.19\textwidth}
      \includegraphics[width=\linewidth]{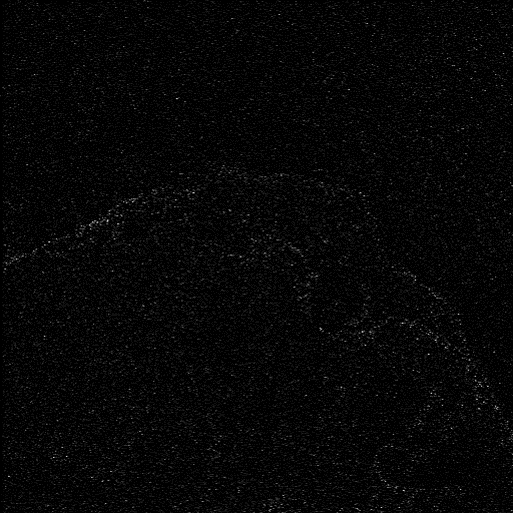}
      \captionsetup{justification=centering}
      \caption{\footnotesize one sub-acquisition, $\lambda/n = 0.125$}
      \label{fig:one_sub_dwt_200_1}
    \end{subfigure} &
    \begin{subfigure}{0.19\textwidth}
      \includegraphics[width=\linewidth]{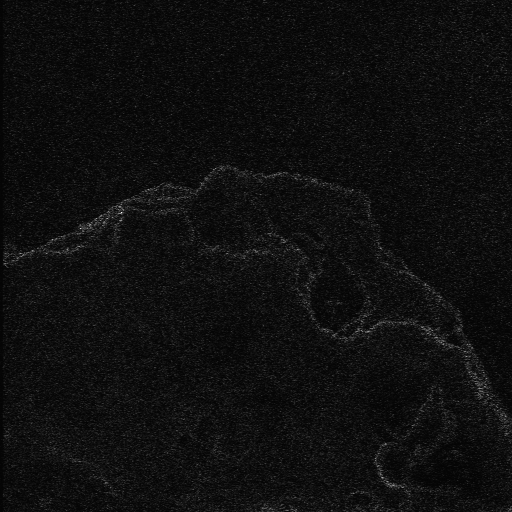}
      \captionsetup{justification=centering}
      \caption{\footnotesize conventional, $\lambda = 1$, $\MSEcaption =15.73$}
      \label{fig:conven_PPNQ_n_8}
    \end{subfigure} &
    \begin{subfigure}{0.19\textwidth}
      \includegraphics[width=\linewidth]{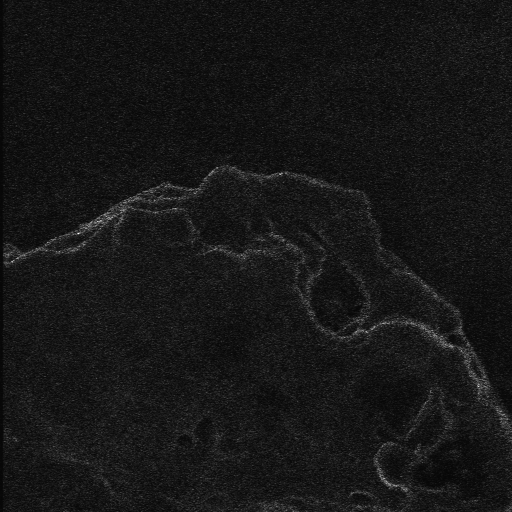}
      \captionsetup{justification=centering}
      \caption{\footnotesize conventional, $\lambda = 2.5$, $\MSEcaption =5.99$}
      \label{fig:conven_PPNQ_n_20}
    \end{subfigure} &
    \begin{subfigure}{0.19\textwidth}
      \includegraphics[width=\linewidth]{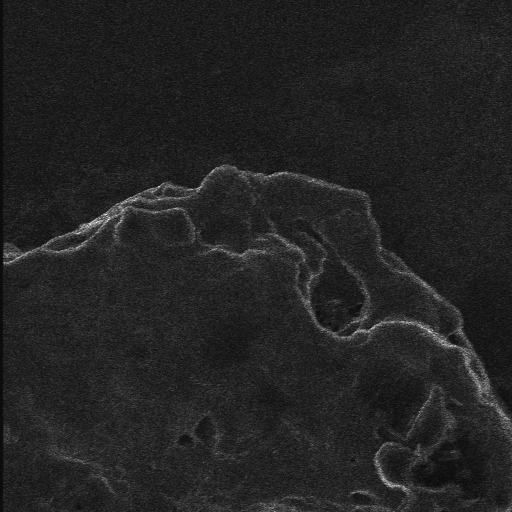}
      \captionsetup{justification=centering}
      \caption{\footnotesize conventional, $\lambda = 7.5$, $\MSEcaption = 1.66$}
      \label{fig:conven_PPNQ_n_60}
    \end{subfigure} &
    \begin{subfigure}{0.19\textwidth}
      \includegraphics[width=\linewidth]{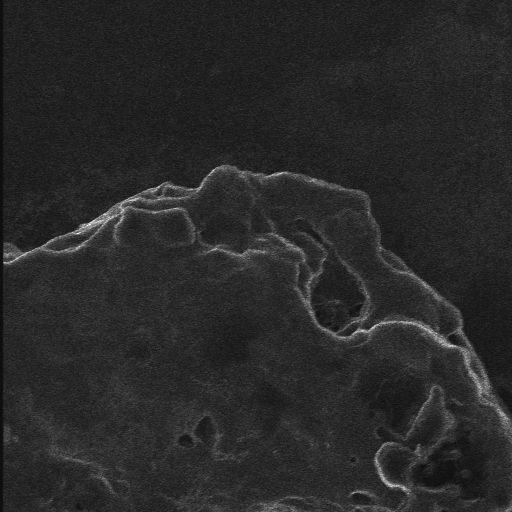}
      \captionsetup{justification=centering}
      \caption{\footnotesize conventional, $\lambda = 16$, $\MSEcaption = 0.51$}
      \label{fig:conven_PPNQ_n_128}
    \end{subfigure} \\
    \begin{subfigure}[t]{0.19\textwidth}
      \includegraphics[width=\linewidth]{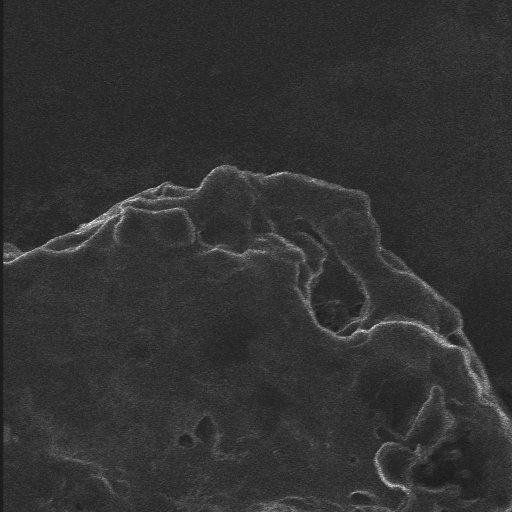}
      \captionsetup{justification=centering}
      \caption{\footnotesize ground truth proxy}
      \label{fig:GT_real_data_PPNQ}
    \end{subfigure} &
    \begin{subfigure}[t]{0.19\textwidth}
      \includegraphics[width=\linewidth]{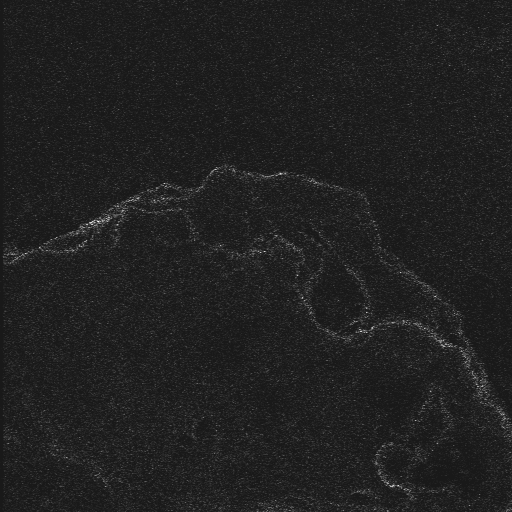}
      \captionsetup{justification=centering}
      \caption{\footnotesize time-resolved, $\lambda = 1$, $\MSEcaption = 3.82$}
      \label{fig:TR_real_data_PPNQ_n8}
    \end{subfigure} &
    \begin{subfigure}[t]{0.19\textwidth}
      \includegraphics[width=\linewidth]{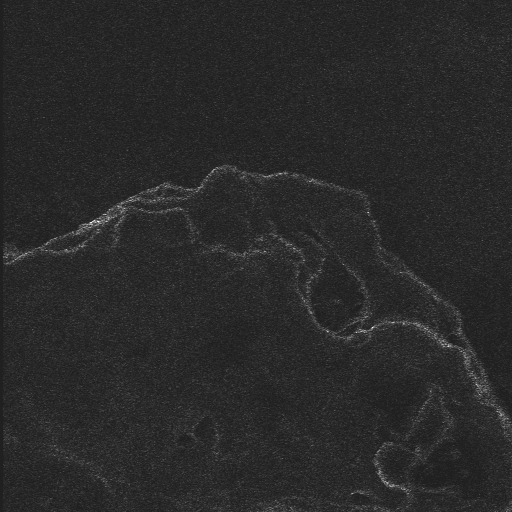}
      \captionsetup{justification=centering}
      \caption{\footnotesize time-resolved, $\lambda = 2.5$, $\MSEcaption = 1.63$}
      \label{fig:TR_real_data_PPNQ_n20}
    \end{subfigure} &
    \begin{subfigure}[t]{0.19\textwidth}
      \includegraphics[width=\linewidth]{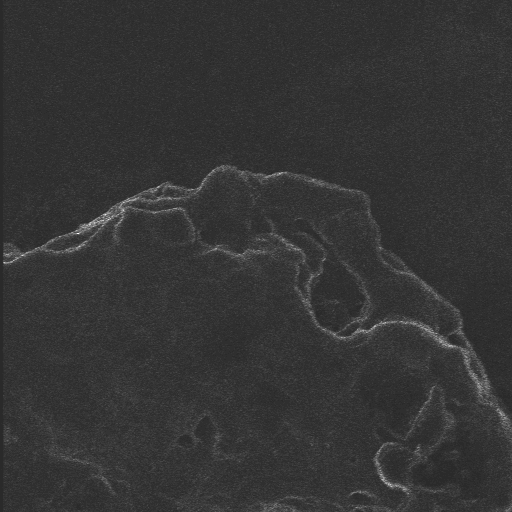}
      \captionsetup{justification=centering}
      \caption{\footnotesize time-resolved, $\lambda = 7.5$, $\MSEcaption = 0.78$}
      \label{fig:TR_real_data_PPNQ_n60}
    \end{subfigure} &
    \begin{subfigure}[t]{0.19\textwidth}
      \includegraphics[width=\linewidth]{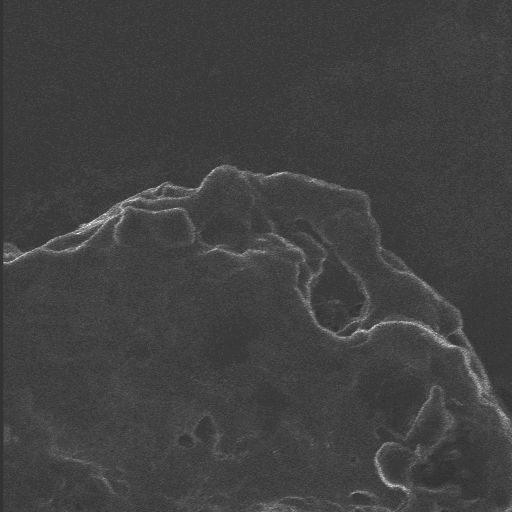}
      \captionsetup{justification=centering}
      \caption{\footnotesize time-resolved, $\lambda = 16$, $\MSEcaption = 0.51$}
      \label{fig:TR_real_data_PPNQ_n_128}
    \end{subfigure} 
  \end{tabular}
  \caption{HIM experimental results for imaging a carbon-based defect on a silicon substrate sample.
  Results for our time-resolved method use the Quantized Poisson--Poisson--Normal compound model in Section~\ref{subsec:PPN_quantized_Model} with $c_1 = 5$ and $c_2 = 50$.
  All images are produced pixel-by-pixel (i.e., without spatial regularization).
  (a) A typical image from one sub-acquisition acquired with dose
  $\lambda/n = 0.125$ ions per pixel.
  (b) Conventional method using 8 sub-acquisitions.
  (c) Conventional method using 20 sub-acquisitions.
  (d) Conventional method using 60 sub-acquisitions.
  (e) Conventional method using 128 sub-acquisitions.
  (f) Ground truth proxy with dose $\lambda = 16$, formed by averaging results shown in (e) and (j).
  (g) Time-resolved method using 8 sub-acquisitions.
  (h) Time-resolved method using 20 sub-acquisitions.
  (i) Time-resolved method using 60 sub-acquisitions.
  (j) Time-resolved method using 128 sub-acquisitions.
  Comparing (b) and (f) shows MSE reduction by a factor of 3.67 from our time-resolved method,
  and comparing (c) and (f) shows dose reduction by a factor of 3.0 without increase in MSE\@.
  }
  \label{fig:real_data_PPNQ_all}
\end{figure*}

Our methods were validated with data from a Zeiss ORION NanoFab HIM used to image
a carbon-based defect on a silicon substrate.
The instrument was used to collect 128 sub-acquisitions of the sample
using a 0.1~pA beam current and 200~ns dwell time,
resulting in low ion dose of 0.125 ions per pixel.
The image of one typical sub-acquisition is shown in Figure~\ref{fig:one_sub_dwt_200_1}.
In this and all other panels of Figure~\ref{fig:real_data_PPNQ_all},
the scaling for display maps the range of the data linearly to the full
black-to-white range.

With the set of 128 sub-acquisitions, we can emulate conventional
and time-resolved image formation for doses from 0.125 ions per pixel
to 16 ions per pixel.
Conventional image formation has no time resolution;
this is emulated by summing the sub-acquisitions, as shown in
Figures~\ref{fig:conven_PPNQ_n_20}, \ref{fig:conven_PPNQ_n_60},
and~\ref{fig:conven_PPNQ_n_128}.
For our time-resolved method,
since the instrument does not use direct electron detection and its output at each pixel is a nonnegative integer,
the Quantized Poisson--Poisson--Normal model of Section~\ref{subsec:PPN_quantized_Model} was employed.
Hyper-parameters $c_1 = 5$ and $c_2 = 50$ were used without significant optimization.
Results of pixel-by-pixel ML estimation under this model are shown in Figures~\ref{fig:TR_real_data_PPNQ_n20},
\ref{fig:TR_real_data_PPNQ_n60}, and~\ref{fig:TR_real_data_PPNQ_n_128}.
With increasing ion dose (moving from second to third to fourth column of Figure~\ref{fig:real_data_PPNQ_all}),
the image quality improves as expected.

\subsection{Quantitative evaluation}
\label{sec:quantitative}
With no ground truth image of the sample available,
any accuracy claims are delicate.
We define the \emph{MSE estimate} $\MSE$ for an image as
the average of the squared difference between the image and a proxy for ground truth
(Figure~\ref{fig:GT_real_data_PPNQ})
that is formed by taking the average of the two images produced using the conventional
(Figure~\ref{fig:conven_PPNQ_n_128})
and time-resolved
(Figure~\ref{fig:TR_real_data_PPNQ_n_128})
methods with all 128 sub-acquisitions.
The difference is computed after scaling such that the mean brightness is matched to Figure~\ref{fig:GT_real_data_PPNQ},
on a 0 to 255 scale;%
\footnote{Some consideration of scaling is necessary because the TR method provides estimates of $\eta$ (which usually is in $[2,\, 8]$),
while the conventional estimate is a simple averaging of the output images of the HIM instrument,
after data conversion and processing
for display on a $[0,\, 255]$ scale.}
thus, the units of $\MSE$ are consistent but arbitrary.
These MSE estimates appear in the captions of Figure~\ref{fig:real_data_PPNQ_all}.
The choice of ground truth proxy is open to criticism,
and more conservative quantitative comparisons are discussed in Section~\ref{sec:conservative}.

Comparing Figures~\ref{fig:conven_PPNQ_n_8} and~\ref{fig:TR_real_data_PPNQ_n8} shows
a reduction of $\MSE$ by a factor of 4.12 at very low dose, while comparing Figures~\ref{fig:conven_PPNQ_n_20} and~\ref{fig:TR_real_data_PPNQ_n20} shows
a reduction of $\MSE$ by a factor of 3.67\@.
Similarly, 
comparing Figures~\ref{fig:conven_PPNQ_n_60} and~\ref{fig:TR_real_data_PPNQ_n60} shows
a reduction of $\MSE$ by a factor of 2.13\@.
As discussed further in Section~\ref{sec:conservative},
the reduction in improvement factor in $\MSE$ as dose is
increased is inevitable from the method of computing $\MSE$ and
does not imply that improvement is diminishing.

An alternative is to assert a dose reduction.
Comparing Figures~\ref{fig:conven_PPNQ_n_60} and~\ref{fig:TR_real_data_PPNQ_n20},
the proposed time-resolved measurement method achieves slightly lower $\MSE$ with dose reduced by a factor of 3\@.

\subsection{Conservative error analysis}
\label{sec:conservative}
While we believe $\MSE$ to be a reasonable metric,
it is possible that it presents an inaccurate view of the improvement due to time-resolved sensing.
Thus, we augment the comparison of $\MSE$ values with a decidedly more conservative approach.

\begin{figure}
    \centering
    \includegraphics[width=0.9\linewidth]{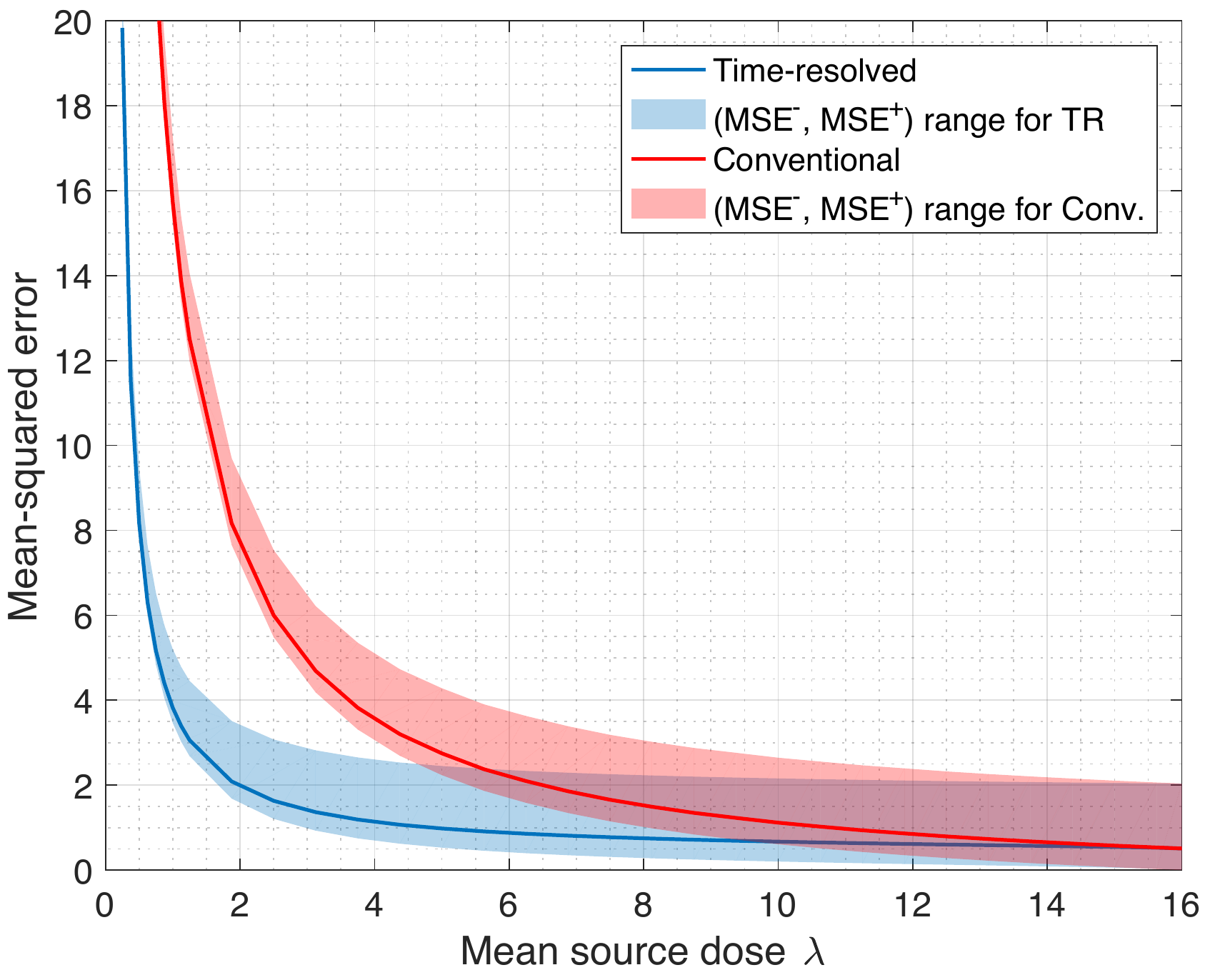}
    \caption{Estimated mean-squared error $\MSEcaption$ (see Section~\ref{sec:quantitative})
       as a function of mean source dose $\lambda$ for conventional (red) and time-resolved sensing (blue) methods.
       Also shown are the ranges $(\MSElowercaption,\,\MSEuppercaption)$ (see Section~\ref{sec:conservative})
       intended to allow more conservative comparisons.}
    \label{fig:MSEcomparison_TRvsConv}
\end{figure}

Accumulating the sequence of 128 sub-acquisitions with conventional image formation creates a sequence of images,
culminating in the $\lambda = 16$ image shown in Figure~\ref{fig:conven_PPNQ_n_128};
similarly, the TR method creates a sequence culminating in Figure~\ref{fig:TR_real_data_PPNQ_n_128}.
Treating Figure~\ref{fig:conven_PPNQ_n_128} as a ground truth proxy would be optimistic for the
sequence of conventionally formed images and thus pessimistic for the sequence of images formed with the TR method.
Conversely,
treating Figure~\ref{fig:TR_real_data_PPNQ_n_128} as a ground truth proxy would be optimistic for the
sequence of images formed with TR method and thus pessimistic for the sequence of conventionally formed images.%
\footnote{Stated differently:
Comparing a conventionally formed image to Figure~\ref{fig:conven_PPNQ_n_128} likely underestimates its error,
while comparing it to Figure~\ref{fig:TR_real_data_PPNQ_n_128} likely overestimates its error.
Conversely,
comparing an image formed with the TR method to Figure~\ref{fig:TR_real_data_PPNQ_n_128} likely underestimates its error,
while comparing it to Figure~\ref{fig:conven_PPNQ_n_128} likely overestimates its error.}
Using Figures~\ref{fig:conven_PPNQ_n_128} and~\ref{fig:TR_real_data_PPNQ_n_128} as ground truth proxies
thus gives an optimistic MSE estimate $\MSElower$ and pessimistic MSE estimate $\MSEupper$ for any image.
These provide a range that is shown along with $\MSE$ in Figure~\ref{fig:MSEcomparison_TRvsConv}.

While $\MSElower$ and $\MSEupper$ are not rigorously lower and upper bounds to the MSE,
they strengthen the evidence that the TR method provides a substantial improvement.
For example, we see that for ion doses up to 4.5,
$\MSEupper$ for the TR method is lower than $\MSElower$ for the conventional method.
Importantly, the convergences of curves for the conventional and TR methods at the maximum ion dose
of $16$
should \emph{not} be construed as showing diminishing advantage for the TR method at higher doses.
As shown in Figure~\ref{fig:MSEcomparison_TRvsConv},
the $\MSElower$ values reach zero at whatever is the highest available ion dose,
the $\MSEupper$ values reach the per-pixel Euclidean distance squared between
Figures~\ref{fig:conven_PPNQ_n_128} and~\ref{fig:TR_real_data_PPNQ_n_128}, and
the $\MSE$ values reach one quarter of that distance.

\section{Discussion}
\label{sec:conclusion}

The main contribution of this paper is to introduce the idea that a set of low-dose focused ion beam microscope measurements can be substantially more informative than a single measurement with the same total dose.
We refer to the acquisition of the set of low-dose measurements as ``time-resolved measurement'' because it can be realized by keeping beam current and total dwell time unchanged, while dividing the dwell time into short time segments.

Our demonstrations of the potential of TR measurements take a few forms.
For a Poisson--Poisson model
(Section~\ref{subsec:PPModel})
that serves as an abstract model for FIB measurement with direct detection of secondary electrons,
we used normalized Fisher information to demonstrate that low-dose measurements are the most informative per incident ion
(Figure~\ref{fig:normalized_Fisher_info})
and yield a substantial multiplicative increase in FI
(Figure~\ref{fig:Fisher_info_PP});
furthermore, we used simulations to demonstrate that ML estimation achieves performance improvement consistent with the FI increase
(Figure~\ref{fig:shep_logan_PP}).
Indirect detection of secondary electrons can be modeled as well
(Section~\ref{sec:hierarchical_models}).
While analysis is made more complicated by these hierarchical models, FI computations and imaging simulations indicate that substantial improvements are still possible
(Figures~\ref{fig:Fisher_info_PPNQ}--\ref{fig:Fisher_info_PPP}).
Experiments with HIM data used a Quantized Poisson--Poisson--Normal model
(Section~\ref{subsec:PPN_quantized_Model}) and
demonstrated the advantage of TR measurements and processing, even without direct electron counting (Figures~\ref{fig:real_data_PPNQ_all} and~\ref{fig:MSEcomparison_TRvsConv}).

TR measurement is not a panacea, and this may become more intuitive by considering settings in which it provides no advantage.
For example, keeping all else unchanged
(aperture, electronic gain, etc.),
when taking a digital photograph
there is no advantage
from dividing some appropriate exposure time into 100 shorter exposures.
Since the original exposure time does not cause saturation,
the 100 shorter-exposure photographs should simply be added together.
If anything, the 100 shorter exposures is worse because
each frame is subject to readout noise.

For a more formal demonstration,
suppose i.i.d.\ Poisson random variables
$X_1,\,X_2,\,\ldots,\,X_n$ are observed
in analogy to time-resolved measurement,
with observation of only $X = \sum_{i = 1}^{n} X_i$ as the
counterpart \emph{without} time resolution.
If each $X_i$ has mean parameter $\lambda' = \lambda/n$,
then $X$ is a Poisson random variable with mean parameter $\lambda$.
There is no FI difference between
$(X_1,\,X_2,\,\ldots,\,X_n)$ and $X$
when the goal is estimation of $\lambda$,
so time-resolved sensing does not provide any advantage in this case:
\begin{align*}
\mathcal{I}_{X_1,X_2,\ldots,X_n}(\lambda)
 &= n \, \mathcal{I}_{X_1}(\lambda)
  \eqlabel{a} n \, \frac{1}{n^2} \mathcal{I}_{X_1}(\lambda')
  \eqlabel{b} \frac{1}{n} \, \frac{1}{\lambda'} \\
 &= \frac{1}{n} \, \frac{1}{\lambda/n}
  = \frac{1}{\lambda}
  \eqlabel{c} \mathcal{I}_{X}(\lambda),
\end{align*}
where (a) follows from the reparameterization rule for FI~\cite[(13.21)]{Lehman1998}; and
(b) and (c) from the FI of Poisson parameter $\theta$ being $1/\theta$.
For another example, suppose
$X_1,\,X_2,\,\ldots,\,X_n$  are i.i.d.\ Bernoulli random variables with parameter $p$.
Then 
$X = \sum_{i = 1}^{n} X_i$ is a binomial($n$,$p$) random variable,
and well-known FI expressions give
\begin{equation}
\mathcal{I}_{X_1,X_2,\ldots,X_n}(p)
= n \, \mathcal{I}_{X_1}(p)
= \frac{n}{p(1-p)}
= \mathcal{I}_{X}(p).
\end{equation}
It is the compound nature of FIB microscopy measurements creates the potential for improvement from TR sensing.

\appendix
\section{Neyman Type A distribution of the number of secondary electrons}
\label{app:NeymanTypeA}
We wish to derive the PMF of $Y$ in \eqref{equ:Y-as-sum},
where $M \sim \Poisson(\lambda)$ and $X_i \sim \Poisson(\eta)$ for each $i$.
Since the sum of a deterministic number of Poisson random variables is a Poisson random variable,
given $M = m$, $Y$ is a Poisson random variable with mean
$m\eta$.
The
PMF of $Y$ can now be derived by marginalizing the joint PMF of $Y$ and $M$ over $M$:
\begin{align*}
P_Y(y)
&= \sum\limits_{m = 0}^{\infty} P_{Y,M}(y,m) 
 \eqlabel{a} \sum\limits_{m = 0}^{\infty} P_{Y|M}(y \smid m) \, P_M(m) \\
&\eqlabel{b} \sum\limits_{m = 0}^{\infty} \cfrac{e^{-m \eta} (m\eta)^y}{y!} \frac{e^{-\lambda} \lambda^m}{m!}
 = \cfrac{e^{-\lambda} \eta^y}{y!} \sum\limits_{m = 0}^{\infty} \cfrac{(\lambda e^{-\eta}) ^m m^y}{m!},
\end{align*}
where (a) follows from the multiplication rule; and (b) from substituting Poisson PMFs.
This verifies \eqref{equ:neyman}.
The mean in \eqref{eq:Y-mean} and variance in \eqref{eq:Y-variance}
follow from the laws of total expectation and of total variance,
each applied with conditioning on $M$.

\section{Derivation of Fisher information under Neyman Type A model}
\label{app:FI_PP_expression}
For derivation of \eqref{equ:Fisher_info_equation_PP},
let us first write $\log P_Y(y \sMid \eta, \lambda)$
using \eqref{equ:neyman}:
\begin{equation*}
\log P_Y(y \sMid \eta, \lambda)
= -\lambda + y\log \eta - \log y! 
  + \log\Bigg( \sum\limits_{m = 0}^{\infty}\frac{(\lambda e^{-\eta})^m m^y}{m!} \Bigg).
\end{equation*}
Taking the derivative with respect to $\eta$, we find
\begin{eqnarray*}
\lefteqn{ \frac{\partial \log P_Y(y \sMid \eta, \lambda)}{\partial \eta} 
 = \frac{y}{\eta} - \frac{\sum\limits_{m = 0}^{\infty} \dfrac{m(\lambda e^{-\eta})^m m^y}{m!}}{\sum\limits_{m = 0}^{\infty} \dfrac{(\lambda e^{-\eta})^m m^y}{m!}} } \\
&=& \frac{y}{\eta} - \frac{\sum\limits_{m = 0}^{\infty} \dfrac{(\lambda e^{-\eta})^m m^{y + 1}}{m!}}{\sum\limits_{m = 0}^{\infty} \dfrac{(\lambda e^{-\eta})^m m^y}{m!}} \\
& \eqlabel{a} & \frac{y}{\eta}
    - \frac{{P_Y(y + 1 \sMid \eta, \lambda)}\bigg/{\dfrac{e^{-\lambda} \eta^{y + 1}}{(y + 1)!}}}
           {{P_Y(y \sMid \eta, \lambda)}\bigg/{\dfrac{e^{-\lambda} \eta^{y}}{y!}}} \\
&=& \frac{y}{\eta} - \frac{P_Y(y + 1 \sMid \eta, \lambda)}{P_Y(y \sMid \eta, \lambda)} \frac{y + 1}{\eta}, \\
\end{eqnarray*}
where (a) follows from \eqref{equ:neyman}.
Then the Fisher information is the second moment of the above expression, which verifies
\eqref{equ:Fisher_info_equation_PP}.

\section{Normalized Fisher information limits under Neyman Type A model}
\label{app:FI_Neyman_limits}

\subsection{Low-dose limit}
To evaluate
$\lim_{\lambda \rightarrow 0} \Frac{\mathcal{I}_Y(\eta \sMid \lambda)}{\lambda}$,
we first find $\lambda \rightarrow 0$ limits of expressions that appear in
\eqref{equ:Fisher_info_equation_PP}, including both the
PMF in \eqref{equ:neyman} and the probability ratio
$\Frac{P_Y(y+1 \sMid \eta, \lambda )}
      {P_Y(y \sMid \eta, \lambda )}$.

For $y = 0$,
\begin{align}
P_Y(0 \sMid \eta, \lambda)
&= \frac{e^{-\lambda} \eta^0}{0!} \sum\limits_{m = 0}^{\infty} \frac{(\lambda e^{-\eta})^m m^0}{m!} \nonumber \\
&\eqlabel{a} e^{-\lambda} \sum\limits_{m = 0}^{\infty} \frac{(\lambda e^{-\eta})^m}{m!} 
 \eqlabel{b} e^{-\lambda}\exp(\lambda e^{-\eta}),
 \label{eq:P_Y_0}
\end{align}
where (a) follows from $m^0 =1$; and
(b) from identifying the series expansion of the exponential function.
Similarly, for $y = 1$,
\begin{align}
P_Y(1 \sMid \eta, \lambda)
&= \frac{e^{-\lambda} \eta^1}{1!} \sum_{m = 0}^{\infty} \frac{(\lambda e^{-\eta})^m m^1}{m!} \nonumber \\
&= (e^{-\lambda} \eta)(\lambda e^{-\eta})\exp(\lambda e^{-\eta}),
 \label{eq:P_Y_1}
\end{align}
and for $y = 2$,
\begin{align}
P_Y(2 \sMid \eta, \lambda)
&= \frac{e^{-\lambda} \eta^2}{2!} \sum_{m = 0}^{\infty} \frac{(\lambda e^{-\eta})^m m^2}{m!} \nonumber \\
&= \frac{e^{-\lambda} \eta^2}{2}
(\lambda e^{-\eta})
(1 + \lambda e^{-\eta})
\exp(\lambda e^{-\eta}).
 \label{eq:P_Y_2}
\end{align}
For general $y > 0$,
\begin{align}
P_Y(y \sMid \eta, \lambda)
&= \frac{e^{-\lambda} \eta^y}{y!} \sum\limits_{m = 0}^{\infty} \frac{(\lambda e^{-\eta})^m m^y}{m!} \nonumber \\
&= \frac{e^{-\lambda} \eta^y}{y!}
(\lambda e^{-\eta})\,
\mathrm{poly}_{y-1}(\lambda e^{-\eta})
\exp(\lambda e^{-\eta}),
 \label{eq:P_Y_y}
\end{align}
where $\mathrm{poly}_y(\lambda e^{-\eta})$
is a degree-$y$ polynomial in $\lambda e^{-\eta}$
with unit constant term.
This allows us to conclude, for any $y > 0$,
\begin{equation}
\label{eq:lim_P_y_normalized}
  \lim_{\lambda \rightarrow 0} \frac{P_Y(y \sMid \eta, \lambda)}{\lambda}
   = \frac{\eta^y}{y!} e^{-\eta}.
\end{equation}

From \eqref{eq:P_Y_0} and \eqref{eq:P_Y_1}, we obtain, for $y = 0$,
\begin{equation}
\label{eq:P_ratio_0}
\frac{P_Y(y + 1 \sMid \eta, \lambda)}{P_Y(y \sMid \eta, \lambda)}
= \frac{P_Y(1 \sMid \eta, \lambda)}{P_Y(0 \sMid \eta, \lambda)}
= \eta \lambda e^{-\eta}.
\end{equation}
From \eqref{eq:P_Y_1} and \eqref{eq:P_Y_2}, we obtain, for $y = 1$,
\begin{equation}
\label{eq:P_ratio_1}
\frac{P_Y(y + 1 \sMid \eta, \lambda)}{P_Y(y \sMid \eta, \lambda)}
= \frac{P_Y(2 \sMid \eta, \lambda)}{P_Y(1 \sMid \eta, \lambda)}
= \frac{1}{2}\eta (1 + \lambda e^{-\eta}).
\end{equation}
For general $y > 0$, it follows from \eqref{eq:lim_P_y_normalized} that
\begin{equation}
\label{eq:lim_P_ratio}
\lim_{\lambda \rightarrow 0}
  \frac{P_Y(y + 1 \sMid \eta, \lambda)}{P_Y(y \sMid \eta, \lambda)}
  = \frac{\eta}{y + 1}.
\end{equation}

Now to evaluate
$\lim_{\lambda \rightarrow 0} \Frac{\mathcal{I}_Y(\eta \sMid \lambda)}{\lambda}$,
we can pass the limit through to each term in
\eqref{equ:Fisher_info_equation_PP}.
The first term is
\begin{align}
\lim_{\lambda \rightarrow 0}
 & \left(\frac{0}{\eta} - \frac{P_Y(1 \sMid \eta,\lambda)}{P_Y(0 \sMid \eta,\lambda)} \frac{1}{\eta} \right)^2
   \frac{P_Y(0 \sMid \eta,\lambda)}{\lambda} \nonumber \\
 &\eqlabel{a}  \lim_{\lambda \rightarrow 0}
   \left(\frac{0}{\eta} -
   \eta \lambda e^{-\eta}
   \frac{1}{\eta} \right)^2
   \frac{e^{-\lambda}\exp(\lambda e^{-\eta})}{\lambda} \nonumber \\
 &= 0,
\end{align}
where (a) follows from \eqref{eq:P_Y_0} and \eqref{eq:P_ratio_0}.
By substituting \eqref{eq:lim_P_y_normalized} and \eqref{eq:lim_P_ratio}
in \eqref{equ:Fisher_info_equation_PP},
the remaining terms give
\begin{align*}
\lim_{\lambda \rightarrow 0}
 \frac{\mathcal{I}_Y(\eta \sMid \lambda)}{\lambda}
 &= \sum_{y=1}^\infty
        \left(\frac{y}{\eta}
              - \frac{\eta}{y+1} \frac{y+1}{\eta} \right)^2
              \frac{\eta^y}{y!}e^{-\eta} \\
 &= \sum_{y=1}^\infty
        \left(\frac{y}{\eta}
              - 1 \right)^2
              \frac{\eta^y}{y!}e^{-\eta} \\
 &= \left(\frac{e^{\eta}}{\eta} - 1\right) e^{-\eta}
  = \frac{1}{\eta} - e^{-\eta} .
\end{align*}
This proves \eqref{eq:FI-low-lambda}, as desired.

\subsection{High-dose limit}
Let us first compute the Fisher information for
the parameter $\eta$ when a Gaussian random variable
has mean $\eta$ and variance $f(\eta)$ for some
twice-differentiable function $f$.
Let $S \sim \mathcal{N}(\eta,\,f(\eta))$.
Then the log-likelihood of $S$ is
\begin{equation}
\log f_S(s \sMid \eta)
  = - \frac{1}{2}\log (2\pi)
    - \frac{1}{2}\log f(\eta)
    - \frac{(s-\eta)^2}{2f(\eta)}.    
\end{equation}
The derivative of $\log f_S(s \sMid \eta)$ with respect to $\eta$ is
\begin{align*}
\frac{\partial \log f_S(s \sMid \eta)}{\partial \eta}
&= - \frac{f'(\eta)}{2f(\eta)}
   - \frac{2(\eta - s)f(\eta) - (\eta-s)^2 f'(\eta)}{2f(\eta)^2}.
\end{align*}
The second derivative is then
\begin{align*}
\frac{\partial^2 \log f_S(s \sMid \eta)}{\partial \eta^2}
&= - \frac{f''(\eta)f(\eta) - f'(\eta)^2}{2f(\eta)^2}
   - \frac{1}{f(\eta)}
   + \frac{2f'(\eta)}{f(\eta)^2}(\eta - s)\\
&  \quad - \frac{ 2[f'(\eta)]^2 - f''(\eta)f(\eta)}{2f(\eta)^3} (\eta-s)^2.
\end{align*}    
The Fisher information for the estimation of $\eta$ is
\begin{align}
\mathcal{I}_S(\eta)
&= \E{ -\frac{\partial^2 \log f_S(S \sMid \eta)}{\partial \eta^2} \sMid \eta } \nonumber\\
&= \frac{f''(\eta)f(\eta) - f'(\eta)^2}{2f(\eta)^2}
   + \frac{1}{f(\eta)}
   - \frac{2f'(\eta)}{f(\eta)^2}\E{\eta - S} \nonumber\\
&  \quad + \frac{2[f'(\eta)]^2 - f''(\eta)f(\eta)}{2f(\eta)^3} \E{(\eta - S)^2} \nonumber\\
&\eqlabel{a} \frac{f''(\eta)f(\eta) - f'(\eta)^2}{2f(\eta)^2}
   + \frac{1}{f(\eta)}
   - \frac{2f'(\eta)}{f(\eta)^2}\cdot 0 \nonumber\\
&  \quad + \frac{ 2[f'(\eta)]^2 - f''(\eta)f(\eta)}{2f(\eta)^3} \cdot f(n) \nonumber\\
&= \frac{1}{f(\eta)}
   + \frac{[f'(\eta)]^2}{2f(\eta)^2},
\label{eq:FI_S}
\end{align}
where (a) follows from substituting
$\E{\eta-S} = 0$ and $\iE{(\eta-S)^2} = \var{S} = f(\eta)$.
(Note that this simplifies to the familiar
reciprocal of the variance when $f(\eta)$ is a constant.)

At high dose, $Y/\lambda$ is well-approximated as a
$\mathcal{N}(\eta,\,\eta(\eta+1)/\lambda)$ random variable~\cite[Sect.~IV]{Teich:81}.
Thus, define
$f(\eta) = \Frac{\eta(\eta+1)}{\lambda}$
so that $Y/\lambda$ is approximated well by $S$.
Substituting
$f'(\eta) = \Frac{(2\eta+1)}{\lambda}$
into \eqref{eq:FI_S} gives
\begin{align}
\mathcal{I}_S(\eta) 
&= \frac{\lambda}{\eta(\eta+1)}
   + \frac{(2\eta+1)^2}{2\eta^2(\eta+1)^2}.
\label{equ:Fisher_info_PP_large_lambda}
\end{align}
Since $Y \approx \lambda S$,
\[
  \lim_{\lambda \rightarrow \infty}
  \frac{\mathcal{I}_Y( \eta \sMid \lambda)}{\lambda}
  = \lim_{\lambda \rightarrow \infty}
    \left[ \frac{1}{\eta(\eta+1)}
    + \frac{(2\eta+1)^2}{2 \lambda \eta^2(\eta+1)^2} \right]
  = \frac{1}{\eta(\eta+1)},
\]
as desired.

\section*{Declarations of interest}
The authors declare no competing financial interests.

\section*{Author contributions}
KKB and VKG conceived of time-resolved measurement in FIB microscopy.
MP, JMB, and VKG derived the mathematical results.
MP wrote software for image formation and completed all numerical experiments.
MP, JMB, and VKG wrote the manuscript.
All authors edited the manuscript.

\section*{Acknowledgments}
The authors thank John Notte and Deying Xia of Carl Zeiss Microscopy LLC for enlightening discussions and experimental data and images.

Funding:  This material is based upon work supported in part by the US National Science Foundation under Grant No.\ 1422034 and Grant No.\ 1815896.

\bibliographystyle{elsarticle-num}

\bibliography{bibfile}

\end{document}